\documentclass[11pt,a4paper]{article}
\pdfoutput=1
\usepackage{a4wide}
\usepackage{amsmath}
\usepackage{amssymb}
\usepackage{amsfonts}
\usepackage{cite}
\usepackage{color}
\usepackage{adjustbox}
\usepackage{multirow}
\usepackage{pdflscape}
\usepackage{gensymb}
\usepackage{graphicx}
\usepackage{diagbox}
\usepackage{pifont}
\usepackage{bigstrut}
\usepackage[small,bf]{caption}
\usepackage{enumerate}
\usepackage{tabulary}

\newcommand{\dcp}{\delta_{\mathrm{CP}}}

\newcommand{\bo}[1]{\boldsymbol{#1}}
\newcommand{\be}{\begin{equation}}
\newcommand{\ee}{\end{equation}}

\newcommand{\bt}{\bigstrut[t]}
\newcommand{\btb}{\bigstrut}
\newcommand{\bb}{\bigstrut[b]}
\def\nue{{\nu_e}}
\def\anue{{\bar\nu_e}}
\def\numu{{\nu_{\mu}}}
\def\anumu{{\bar\nu_{\mu}}}

\def\anu{{\bar\nu}}

\def\1{\mathbf{1}}
\def\3{\mathbf{3}}
\def\2{\mathbf{2}}

\def\D{\Delta}

\def\th{\theta}

\numberwithin{equation}{section}

\begin{document}

\begin{titlepage}

\vspace*{-15mm}

\begin{flushright}
IP/BBSR/2017-13 \\
SISSA 53/2017/FISI \\ 
IPMU17-0154 \\
IPPP/17/75 \\
\end{flushright}
\vspace*{0.8cm}

\begin{center}

{\bf\Large{Addressing Neutrino Mixing Models with DUNE and T2HK}} \\[10mm]

{\bf Sanjib Kumar Agarwalla}$^{\, a,b,c,}$\footnote{E-mail: \texttt{sanjib@iopb.res.in}}, 
{\bf Sabya Sachi Chatterjee}$^{\, a,b,}$\footnote{E-mail: \texttt{sabya@iopb.res.in}}, \\[1mm]
{\bf S.~T.~Petcov}$^{\, d,e,}$\footnote{Also at Institute of Nuclear Research and Nuclear Energy, Bulgarian Academy of Sciences, 1784 Sofia, Bulgaria.} 
and
{\bf A.~V.~Titov}$^{\, d,f,}$\footnote{E-mail: \texttt{arsenii.titov@durham.ac.uk}} \\
\vspace{8mm}
$^{a}$\,{\it Institute of Physics, Sachivalaya Marg, Sainik School Post, 
Bhubaneswar 751005, India} \\
\vspace{2mm}
$^{b}$\,{\it Homi Bhabha National Institute, Training School Complex, \\ 
Anushakti Nagar, Mumbai 400085, India} \\
\vspace{2mm}
$^{c}$\,{\it International Centre for Theoretical Physics, 
Strada Costiera 11, 34151 Trieste, Italy} \\
\vspace{2mm}
$^{d}$\,{\it SISSA/INFN, Via Bonomea 265, 34136 Trieste, Italy} \\
\vspace{2mm}
$^{e}$\,{\it Kavli IPMU (WPI), University of Tokyo, 
5-1-5 Kashiwanoha, 277-8583 Kashiwa, Japan} \\
\vspace{2mm}
$^{f}$\,{\it Institute for Particle Physics Phenomenology, 
Department of Physics, Durham University,\\ 
South Road, Durham DH1 3LE, United Kingdom}
\end{center}
\vspace{8mm}

\begin{abstract}
\noindent 
We consider schemes of neutrino mixing arising within the 
discrete symmetry approach to the well-known flavour problem.
We concentrate on $3\nu$ mixing schemes in which the cosine of the 
Dirac CP violation phase $\dcp$ satisfies a sum rule by which 
it is expressed in terms of three neutrino mixing angles $\theta_{12}$, 
$\theta_{23}$, and  $\theta_{13}$, and a fixed real angle 
$\th^\nu_{12}$, whose value depends on the employed discrete 
symmetry and its breaking. 
We consider five underlying symmetry forms 
of the neutrino mixing matrix:
bimaximal (BM), tri-bimaximal (TBM), golden ratio A (GRA) and B (GRB), 
and hexagonal (HG). For each symmetry form, the sum rule 
yields specific prediction for $\cos\dcp$ for 
fixed $\th_{12}$, $\th_{23}$, and $\th_{13}$. 
In the context of the proposed DUNE and T2HK facilities, we study 
(i) the compatibility of these predictions with present neutrino oscillation data, and 
(ii) the potential of these experiments to discriminate between various symmetry forms.
\end{abstract}
\end{titlepage}

\setcounter{footnote}{0}

\section{Introduction and Motivation}
\label{sec:intro}

 All compelling neutrino oscillation data are compatible with 
$3\nu$ mixing \cite{Olive:2016xmw}, i.e., with existence 
of three light neutrino states $\nu_{1,2,3}$ with definite masses 
$m_{1,2,3}$, three orthogonal linear combinations of which  
form the three flavour neutrino states $\nu_e$, $\nu_\mu$ and $\nu_\tau$.
The flavour neutrino (flavour antineutrino) 
oscillation probabilities in the case of $3\nu$ mixing 
are characterised, as is well known, by six fundamental parameters.
In the standard parametrisation of the 
Pontecorvo, Maki, Nakagawa, Sakata (PMNS) 
neutrino mixing matrix \cite{Olive:2016xmw}, these are 
the solar, reactor, and atmospheric mixing angles, 
$\th_{12}$, $\th_{13}$, and $\th_{23}$, respectively, 
the Dirac CP violation (CPV) phase, $\dcp$, and the two 
independent mass squared differences, 
$\Delta m_{21}^2 \equiv m_2^2 - m_1^2$ and 
$\Delta m_{31}^2 \equiv m_3^2 - m_1^2$.  
The mixing angles $\th_{12}$ and $\th_{13}$ 
as well as the solar and the absolute value of the atmospheric 
neutrino mass squared differences, $\Delta m_{21}^2$ and $|\Delta m_{31}^2|$, respectively, 
have been measured in neutrino oscillation experiments with a relatively high precision 
\cite{Esteban:2016qun,Capozzi:2017ipn,deSalas:2017kay}. 
The precision on $\th_{23}$ is somewhat worse, 
the relative $1\sigma$ uncertainty 
on $\sin^2\theta_{23}$ being approximately 10\%.
At the same time, the octant of $\th_{23}$ and 
the value of $\dcp$ remain unknown. 
The sign of $\Delta m_{31}^2$, which is also undetermined, 
allows, as is well known,
for two possible types of 
the neutrino mass spectrum: (i) with normal ordering (NO) 
if $\Delta m_{31}^2 > 0$, and 
(ii) with inverted ordering (IO) if $\Delta m_{31}^2 < 0$.
In Table~\ref{tab:oscparams}, we summarise the best fit 
values and $1\sigma$, $2\sigma$, $3\sigma$ ranges of the oscillation parameters obtained in 
one of the latest global analysis of the neutrino oscillation 
data~\cite{Capozzi:2017ipn}%
\footnote{Global analyses of the neutrino oscillation data were 
also performed recently in Refs.~\cite{Esteban:2016qun} and \cite{deSalas:2017kay}. 
The best fit values and $1\sigma$, $2\sigma$, $3\sigma$ ranges 
of the parameters obtained 
in these two articles 
almost agree with the findings in Ref.~\cite{Capozzi:2017ipn}.}. 
In the case of IO spectrum, instead of $\Delta m_{31}^2$, 
we use the largest positive mass squared difference 
$\Delta m_{23}^2 \equiv m_2^2 - m_3^2$, which corresponds 
to $\Delta m_{31}^2$ of the NO spectrum.
In particular, we see from Table~\ref{tab:oscparams} that for the NO  
mass spectrum, the values of 
$\dcp \in (31\degree,137\degree)$ are already disfavoured at more than $3\sigma$ 
C.L., while the values of $\dcp$ 
between $180\degree$ and $342\degree$ are allowed at $2\sigma$.
\begin{table}[t]
\centering
\renewcommand{\arraystretch}{2.2}
\begin{adjustbox}{max width=\textwidth}
\begin{tabular}{|l|cccc|} 
\hline\bb
Parameter & Best fit & $1\sigma$ range & $2\sigma$ range & $3\sigma$ range \\ 
\hline
$\dfrac{\sin^2\theta_{12}}{10^{-1}}$ & 2.97 & 2.81--3.14 & 2.65--3.34 & 2.50--3.54 \\
$\dfrac{\sin^2\theta_{13}}{10^{-2}}$~(NO) & 2.15 & 2.08--2.22 & 1.99--2.31 & 1.90--2.40 \\
$\dfrac{\sin^2\theta_{13}}{10^{-2}}$~(IO) & 2.16 & 2.07--2.24 & 1.98--2.33 & 1.90--2.42 \\
$\dfrac{\sin^2\theta_{23}}{10^{-1}}$~(NO) & 4.25 & 4.10--4.46 & 3.95--4.70 & 3.81--6.15 \\
$\dfrac{\sin^2\theta_{23}}{10^{-1}}$~(IO) & 5.89 & 4.17--4.48\,$\oplus$\,5.67--6.05 & 3.99--4.83\,$\oplus$\,5.33--6.21 & 3.84--6.36 \\
$\dcp$~[\degree]~(NO) & $248$ & 212--290 & 180--342 & 0--31\,$\oplus$\,137--360 \\
$\dcp$~[\degree]~(IO) & $236$ & 202--292 & 166--338 & 0--27\,$\oplus$\,124--360 \\
\hline
$\dfrac{\Delta m_{21}^{2}}{10^{-5}~\mathrm{eV}^2}$ & 7.37 & 7.21--7.54 & 7.07--7.73 & 6.93--7.96 \\
$\dfrac{\Delta m_{31}^{2}}{10^{-3}~\mathrm{eV}^2}$~(NO) & 2.56 & 2.53--2.60 & 2.49--2.64 & 2.45--2.69 \\
\bb
$\dfrac{\Delta m_{23}^{2}}{10^{-3}~\mathrm{eV}^2}$~(IO) & 2.54 & 2.51--2.58 & 2.47--2.62 & 2.42--2.66 \\
\hline
\end{tabular}
\end{adjustbox}
\caption{The best fit values and $1\sigma$, $2\sigma$, 3$\sigma$ ranges of the 
neutrino oscillation parameters obtained in the global 
analysis of the neutrino oscillation data 
performed in~\cite{Capozzi:2017ipn}. 
NO (IO) stands for normal (inverted) ordering of the neutrino mass spectrum.}
\label{tab:oscparams}
\end{table}

 Understanding the origin of the patterns of neutrino oscillation parameters  
revealed by the data is one of the most challenging problems in neutrino physics.
It is a part of the more general fundamental problem
in particle physics of understanding the origins of
flavour, i.e., the patterns of quark, charged lepton, and 
neutrino masses, and quark and lepton mixing. 
There exists a possibility that the high-precision measurements of the oscillation parameters 
may shed light on the origin of the observed pattern 
of neutrino mixing and  lepton flavour.  
This would be the case if the observed form of neutrino (and possibly quark) mixing 
were determined by an underlying discrete flavour symmetry.  
One of the most striking features of 
the discrete symmetry approach to neutrino mixing and lepton flavour
(see, e.g., \cite{Altarelli:2010gt,Ishimori:2010au,King:2013eh} for reviews), 
is that it leads to  
(i)~fixed predictions of the values of some of the neutrino mixing angles 
and the Dirac CPV phase $\dcp$, 
and/or 
(ii)~existence of correlations between 
some of the mixing angles  
and/or between the mixing angles and $\dcp$  (see, e.g., \cite{Petcov:2014laa,Girardi:2014faa,Girardi:2015zva,Girardi:2015vha,Girardi:2015rwa,
Marzocca:2013cr,Tanimoto:2015nfa,Ballett:2013wya,
Ge:2011ih,Ge:2011qn,Antusch:2011ic,Hanlon:2013ska}).
These correlations are often referred to as neutrino mixing sum rules%
\footnote{Combining the discrete symmetry approach with 
the idea of generalised CP invariance  
\cite{Branco:1986gr,Feruglio:2012cw,Holthausen:2012dk},
which is a generalisation of the standard CP invariance requirement, 
allows one to obtain predictions also for the Majorana 
CPV phases \cite{Bilenky:1980cx} in the PMNS matrix in the case 
of massive Majorana neutrinos 
(see, e.g., 
\cite{Girardi:2013sza,Ballett:2015wia,Turner:2015uta,Girardi:2016zwz,Lu:2016jit,Penedo:2017vtf}
and references quoted therein).}.
Most importantly, these sum rules can be tested using oscillation data 
\cite{Petcov:2014laa,Girardi:2014faa,Girardi:2015zva,Girardi:2015vha,
Hanlon:2013ska,Ballett:2013wya,Ballett:2014dua,Penedo:2017vtf,M.:2014kca}.

Within the discrete flavour symmetry approach,
the PMNS matrix is predicted to have an 
underlying symmetry form, where $\theta_{12}$,  $\theta_{23}$, and $\theta_{13}$
have values which differ by sub-leading perturbative corrections 
from their respective measured values. 
The approach seems very natural in view of the fact that  
$U_{\rm PMNS} = U_e^{\dagger}\, U_{\nu}$, 
where $U_{e}$ and $U_{\nu}$ are $3\times 3$ unitary matrices  
which diagonalise the charged lepton and 
neutrino mass matrices. Typically (but not universally) 
the matrix  $U_{\nu}$ has a certain symmetry form, 
while the matrix $U_{e}$ provides 
the corrections necessary to bring the symmetry values of
the angles in  $U_{\nu}$ to their experimentally measured values.
A sum rule which relates $\cos\dcp$ with $\th_{12}$, $\th_{23}$, and $\th_{13}$,
arising in this approach,
depends on the underlying symmetry form of the PMNS matrix and 
on the form of the ``correcting'' matrix $U_{e}$.

In the present work, we will concentrate on a particular sum rule for 
$\cos\dcp$ derived in \cite{Petcov:2014laa}, which holds for a rather 
broad class of discrete flavour symmetry models. 
According to this sum rule, $\cos\dcp$
is expressed in terms of the three measured neutrino mixing angles 
and one fixed parameter $\theta^\nu_{12}$ determined by 
the underlying discrete symmetry. This sum rule has the following form:   
\be
\cos\dcp = \frac{\tan\th_{23}}{\sin2\th_{12}\, \sin\th_{13}} 
\left[\cos2\th^\nu_{12} + \left(\sin^2\th_{12} - \cos^2\th^\nu_{12}\right)
\left(1 - \cot^2\th_{23}\,\sin^2\th_{13}\right)\right]\,. 
\label{eq:cosdelta}
\ee
In this study, we consider five widely discussed underlying symmetry forms of 
the neutrino mixing matrix, namely, 
bimaximal (BM) \cite{Petcov:1982ya,Vissani:1997pa,Barger:1998ta,Baltz:1998ey}, 
tri-bimaximal (TBM) \cite{Harrison:2002er,Harrison:2002kp,Xing:2002sw,He:2003rm} 
(see also \cite{Wolfenstein:1978uw}), 
golden ratio type A (GRA) \cite{Datta:2003qg,Kajiyama:2007gx,Everett:2008et}, 
golden ratio type B (GRB) \cite{Rodejohann:2008ir,Adulpravitchai:2009bg}, and 
hexagonal (HG) \cite{Albright:2010ap,Kim:2010zub}. 
Each of these symmetry forms is characterised by a specific value of the 
angle $\th^\nu_{12}$ entering into the sum rule given in eq.~\eqref{eq:cosdelta}. 
Namely, $\th^\nu_{12} = 45\degree$ 
(or $\sin^2\theta^\nu_{12} = 0.5$) for BM; 
$\th^\nu_{12} = \arcsin(1/\sqrt3) \approx 35\degree$ 
(or $\sin^2\theta^\nu_{12} = 1/3$) for TBM; 
$\th^\nu_{12} = \arctan(1/\phi) \approx 32\degree$ 
(or $\sin^2\theta^\nu_{12} \approx 0.276$) for GRA, 
$\phi = (1 + \sqrt5)/2$ being the golden ratio; 
$\th^\nu_{12} = \arccos(\phi/2) = 36\degree$ 
(or $\sin^2\theta^\nu_{12} \approx 0.345$) for GRB; 
and $\th^\nu_{12} = 30\degree$ 
(or $\sin^2\theta^\nu_{12} = 0.25$) for HG.

 First, in eq.~\eqref{eq:cosdelta}, we use the best fit values of the neutrino mixing angles 
assuming NO case from Table~\ref{tab:oscparams} to calculate the best fit value of $\cos\dcp$ for a given symmetry form which has a fixed value of $\th^\nu_{12}$. 
We present the obtained values in Table~\ref{tab:deltaBF}.
\begin{table}
\centering
\renewcommand{\arraystretch}{1.2}
\begin{tabular}{|l|ccc|}
\hline\btb
Symmetry form & $\th^\nu_{12}$ [\degree] & $\cos\dcp$ & $\dcp$ [\degree] \\
\hline\bt
BM & $45$ & unphysical & unphysical \\
TBM & $\arcsin(1/\sqrt{3})\approx35$ & $-0.16$ & $\phantom{1}99 \lor 261$ \\
GRA & $\arctan(1/\phi)\approx32$ & $\phantom{-}0.21$ & $\phantom{1}78 \lor 282$ \\
GRB & $\arccos(\phi/2)=36$ & $-0.24$ & $104 \lor 256$ \\
HG & $30$ & $\phantom{-}0.39$ & $\phantom{1}67 \lor 293$ \\
\hline
\end{tabular}
\caption{The best fit values of $\cos\dcp$ and $\dcp$ 
from the sum rule in eq.~\eqref{eq:cosdelta} 
for the different symmetry forms. 
The mixing angles $\th_{12}$,\,$\th_{23}$,\,and $\th_{13}$ 
have been fixed to their NO best fit values from Table~\ref{tab:oscparams}. 
The $\phi$ stands for 
the golden ratio: $\phi = (1+\sqrt{5})/2$. 
See text for further details.}
\label{tab:deltaBF}
\end{table}
For each symmetry form, the predicted value of $\cos\dcp$ gives rise to two 
values of $\dcp$ located symmetrically with respect to zero, 
which are also given in  Table~\ref{tab:deltaBF}. 
Further, we calculate errors on these values by varying 
$\th_{12}$, $\th_{23}$, and $\th_{13}$ (one at a time) in their $3\sigma$ 
experimentally allowed ranges for NO as given in Table~\ref{tab:oscparams}
and fixing the two remaining angles to their best fit values. 
We summarise the obtained intervals of values of $\dcp$ in 
Table~\ref{tab:deltaINTERVALS}.
\begin{table}
\centering
\renewcommand{\arraystretch}{1.2}
{
\newcommand{\mc}[3]{\multicolumn{#1}{#2}{#3}}
\newcommand{\mr}[3]{\multirow{#1}{#2}{#3}}
\begin{tabular}{|l|ccc|}
\hline
\mr{2}{*}{Symmetry form} & 
\mc{3}{c|}{Intervals for $\dcp$ [\degree] obtained varying} \\
& $\th_{12}$ in $3\sigma$ & $\th_{23}$ in $3\sigma$ & $\th_{13}$ in $3\sigma$ \\
\hline\bt
BM & 150--180\,$\lor$\,180--210 
        & unphysical
        & unphysical \\
TBM & \phantom{1}79--119\,$\lor$\,241--281
          & \phantom{1}98--107\,$\lor$\,253--262
          & 98--101\,$\lor$\,259--262\phantom{1} \\
GRA & \phantom{11}57--95\,$\lor$\,265--303 
          & \phantom{11}76--78\,$\lor$\,282--284
          & 77.6--77.9\,$\lor$\,282.1--282.4 \\
GRB & \phantom{1}84--125\,$\lor$\,235--276 
          & 102--114\,$\lor$\,246--258
          & 103--106\,$\lor$\,254--257\phantom{11} \\
HG & \phantom{11}45--84\,$\lor$\,276--315 
       & \phantom{11}60--68\,$\lor$\,292--300
       & \phantom{11}66--68\,$\lor$\,292--294\phantom{11} \\
\hline
\end{tabular}
}
\caption{The intervals for $\dcp$ due to the present $3\sigma$ uncertainties 
in the values of the neutrino mixing angles. 
The quoted intervals are obtained varying one mixing angle 
in its corresponding $3\sigma$ range for the NO spectrum 
and fixing the other two angles to their NO best fit values.}
\label{tab:deltaINTERVALS}
\end{table}

In the case of the BM symmetry form, the 
obtained best fit value of $\cos\dcp = -1.26$ is unphysical. 
This reflects the fact that the BM symmetry form 
does not provide a good description of the present best fit values of 
the neutrino mixing angles, as discussed in \cite{Marzocca:2013cr}.
As can be seen from Table~\ref{tab:deltaINTERVALS}, 
current uncertainties on the mixing angles allow us 
to accommodate physical values of $\cos\dcp$ 
for the BM symmetry form. 
For instance, fixing $\sin^2\th_{13}$ and $\sin^2\th_{23}$ 
to their best fit values, $\cos\dcp = -1$ requires $\sin^2\th_{12} = 0.3343$, 
which is the upper bound of the corresponding $2\sigma$ allowed range 
of  $\sin^2\th_{12}$ (see Table~\ref{tab:oscparams}).

A rather detailed analysis 
of the predictions for $\cos\dcp$ of the sum rule 
in eq.~\eqref{eq:cosdelta} has been performed
in  Refs.~\cite{Girardi:2014faa,Girardi:2015zva}.  
In particular, likelihood profiles for 
$\cos\dcp$ for each symmetry form have been presented  
using the current and prospective precision on the 
neutrino mixing parameters (see Figs.~12 and 13 in \cite{Girardi:2014faa}). 
In the present work, using the potential of the future long-baseline (LBL) 
neutrino oscillation experiments%
\footnote{Recently, the authors of 
Refs.~\cite{Ballett:2016yod,Chatterjee:2017xkb,Chatterjee:2017ilf,Pasquini:2017hby} investigated the capabilities of current and 
future LBL experiments to probe few flavour models, which lead to 
\textit{alternative correlations}
between the neutrino mixing parameters, which 
differ from those considered by us.}, 
namely, Deep Underground Neutrino Experiment (DUNE) and 
Tokai to Hyper-Kamiokande (T2HK), we study in detail 
(i)~to what degree the sum rule predictions for $\cos\dcp$ are compatible with 
the present neutrino oscillation data, and 
(ii)~how well the considered symmetry forms, BM, TBM, GRA, GRB, and HG, 
can be discriminated from each other. 
 
The layout of the article is as follows. In Section~\ref{sec:first-glance}, 
we take a first glance at the sum rule predictions. 
In Section~\ref{sec:set-ups-and-rates}, we give a short 
description of the planned DUNE and T2HK experiments and provide 
expected event rates for both the set-ups. 
Section~\ref{sec:analysis} contains details of the statistical analysis.
In Section~\ref{sec:results}, we present and discuss results of 
this analysis. More specifically, in subsection~\ref{sec:compatibility}, 
we test the compatibility between the considered symmetry 
forms and present oscillation data. 
Next, in subsection~\ref{sec:distinguishing}, we explore the potential of 
DUNE, T2HK, and their combination to distinguish between the symmetry forms 
in question under the assumption that one of them is realised in Nature. 
In subsection~\ref{sec:BM}, we consider the BM symmetry form using 
the values of the mixing angles for which this form is viable, and study 
at which C.L. it can be distinguished from the other symmetry forms 
considered. 
We conclude in Section~\ref{sec:conclusions}. 
Appendix~\ref{app:priors} discusses the issue of 
external priors on $\sin^2\th_{12}$ and $\sin^2\th_{13}$.
In Appendix~\ref{app:dm31sq-marginalisation}, we show
the impact of marginalisation over $\Delta m_{31}^{2}$.
Finally, in Appendix~\ref{app:ss23delta}, we study the compatibility of 
the considered symmetry forms with any potentially true values of 
$\sin^2\th_{23}$ and $\dcp$ in the context of DUNE and T2HK.

\section{A First Glance at the Sum Rule Predictions}
\label{sec:first-glance}

 Let us first take a closer look at the sum rule predictions for $\dcp$ 
summarised in Tables~\ref{tab:deltaBF} and \ref{tab:deltaINTERVALS}. 
Several important points that can be made from these two tables and 
the current global data (see Table~\ref{tab:oscparams}) are in order.
\begin{itemize}
\item The parameter $\th^\nu_{12}$  has a fixed value for each symmetry form as given in Table~\ref{tab:deltaBF}. 
For fixed choices of $\th_{12}$, $\th_{23}$, $\th_{13}$, and $\th^\nu_{12}$, 
eq.~\eqref{eq:cosdelta} predicts a certain value of  
$\cos\dcp$, which gives rise to two values of $\dcp$ 
(see fourth column of Table~\ref{tab:deltaBF}). 
\item The complementarity \cite{Pascoli:2013wca} between the modern reactor 
(Daya Bay, RENO, and Double Chooz) and accelerator (T2K and NO$\nu$A) data
has enabled us to probe the parameter space for $\dcp$. Already, 
the latest global data have disfavoured values of 
$\dcp \in (31\degree,137\degree)$ at $3\sigma$ C.L. 
for NO (see Table~\ref{tab:oscparams}). 
From Table~\ref{tab:deltaINTERVALS}, we see that 
out of the three mixing angles, 
the $3\sigma$ allowed range of $\th_{12}$ causes 
the largest uncertainty in $\dcp$ predicted by eq.~\eqref{eq:cosdelta}.
But, note that, all the symmetry forms except BM 
predict one of the ranges of $\dcp$ in the interval of 
$31\degree$ to $137\degree$, 
which has been already ruled out at $3\sigma$.
Therefore, we will not consider these ranges further in our study, 
and we will only consider the values of $\dcp$ in the interval of 
$180\degree$ to $360\degree$.
\item Further, we notice from Tables~\ref{tab:oscparams} and \ref{tab:deltaINTERVALS} 
that for TBM and GRB, the predicted 
intervals of $\dcp$ lie within the $1\sigma$ experimentally allowed range. 
For BM, GRA, and HG, the intervals of interest fall within the $2\sigma$ range.
Now, it would be quite interesting to assess
the sensitivity of the future LBL experiments 
in discriminating various symmetry forms, 
which is the main thrust of the present work.
\item T2HK and DUNE will not be able to constrain $\th_{12}$, 
which causes the largest uncertainty in predicting the range of $\dcp$ 
(see Table~\ref{tab:deltaINTERVALS}). 
However, the proposed JUNO experiment will provide 
a high-precision measurement of $\sin^2\th_{12}$
with a relative $1\sigma$ uncertainty of 0.7\%.
Therefore, we impose a prior on $\sin^2\th_{12}$ expected from JUNO, 
which we will discuss later in detail in Section~\ref{sec:analysis} and 
in Appendix~\ref{app:priors}.
\item Table~\ref{tab:deltaINTERVALS} shows that the sum rule predictions 
depend to some extent on $\th_{23}$. 
Therefore, we vary this angle both 
in data and in fit. 
The LBL experiments themselves are sensitive to $\th_{23}$.  
Thus, we do not impose any external prior on this angle 
(see details in Section~\ref{sec:analysis}).
\item The experiments under discussion are sensitive to $\th_{13}$ 
through the appearance channels $\numu \to \nue$ and $\anumu \to \anue$. 
Therefore, the role of an external prior on $\sin^2\th_{13}$ is negligible 
for the physics case under study, 
as we will see later in Fig.~\ref{fig:priors}.
However, we put a prior on $\sin^2\th_{13}$ as expected from Daya Bay 
to speed up our simulations (see Section~\ref{sec:analysis} for details).
\end{itemize}

\section{Experimental Features and Event Rates}
\label{sec:set-ups-and-rates}

In this section, we first briefly describe the key experimental features 
of the proposed high-precision DUNE and T2HK facilities that we use
in our numerical simulation.

\subsection{The Next Generation Experiments: DUNE and T2HK}
\label{subsec:set-ups}

The planned Deep Underground Neutrino Experiment (DUNE) aims to
achieve new milestones in the intensity frontier with a new, 
high-intensity, on-axis, wide-band neutrino beam from Fermilab 
directed towards a massive liquid argon time-projection chamber 
(LArTPC) far detector housed at the Homestake Mine in South Dakota 
over a baseline of 1300 km~\cite{Acciarri:2016crz,Acciarri:2015uup,Strait:2016mof,Acciarri:2016ooe,Adams:2013qkq}.
In our simulation, we consider a fiducial mass of 35 kt for the far detector
and the detector characteristics have been taken from Table~1 of 
Ref.~\cite{Agarwalla:2011hh}. As far as beam specifications are
concerned, we assume a modest proton beam power of 708~kW 
in its initial phase with 120~GeV proton energy, which can supply 
$6 \times 10^{20}$ protons on target (p.o.t.) in 188 days per calendar year.
In our calculation, we have used the fluxes which were generated 
assuming a decay pipe length of 200~m and 200~kA horn current~\cite{mbishai}.
We assume that DUNE will collect data for ten years (5 years in $\nu$ mode
and 5 years in $\bar\nu$ mode), which is equivalent to a total exposure 
of 248~$\mathrm{kt} \cdot \mathrm{MW} \cdot \mathrm{year}$\footnote{Note that, our assumptions on
various components of the DUNE set-up differ slightly in comparison 
to the reference design in the Conceptual Design Report (CDR)
of DUNE~\cite{Acciarri:2015uup}. However, it is expected that the reference 
design of the DUNE experiment is going to evolve with time as we will learn
more about this set-up with the help of ongoing R\&D studies.}.
In our simulation, we consider the reconstructed $\nu$ energy range 
to be 0.5 GeV to 10 GeV for both appearance and disappearance 
channels. We take the same energy range for antineutrino as well.

The proposed Hyper-Kamiokande (HK) water Cherenkov detector will 
serve as the far detector of a long-baseline neutrino experiment using 
an upgraded neutrino beam from the J-PARC facility, commonly known 
as ``T2HK'' (Tokai to Hyper-Kamiokande) 
experiment~\cite{Abe:2011ts,Abe:2014oxa,Abe:2015zbg}. 
This set-up is highly sensitive to the Dirac CPV phase $\dcp$ 
of the PMNS $3\nu$ mixing matrix and holds promise to resolve 
the mystery of leptonic CP violation in neutrino 
oscillations at an unprecedented confidence level~\cite{Abe:2014oxa}.
We perform the simulation for T2HK according to 
Refs.~\cite{Abe:2014oxa,Abe:2015zbg}.
To produce an intense $\nu$/$\anu$ beam for HK, we consider an 
integrated proton beam power of 7.5 MW $\times$ $10^7$ seconds, 
which can deliver in total $15.6 \times 10^{21}$ p.o.t. with 
a 30~GeV proton beam. We assume that these total p.o.t. will be 
shared among $\nu$ and $\anu$ modes with a run-time
ratio of 1:3 to have almost equal statistics in both the modes.
The huge 560~kt (fiducial) HK detector will be placed in the 
Tochibora mine, at a distance of 295~km from J-PARC at an
off-axis angle of $\sim 2.5\degree$, which will produce a
narrow band beam with a sharp peak around the first 
oscillation maximum of 0.6~GeV. The total exposure 
that we consider for T2HK is 
4200~$\mathrm{kt} \cdot \mathrm{MW} \cdot \mathrm{year}$.
In our simulation, we take the reconstructed  
$\nu_e$ and $\anue$ energy range of 0.1~GeV to 1.25~GeV 
for the appearance channel. As far as the disappearance channel 
is concerned, the assumed energy range is 0.1~GeV to 7~GeV 
for both $\numu$ and $\anumu$ candidate events.

Recently, the baseline design for T2HK has been revised~\cite{Abe:2016ero}.
According to this latest publication~\cite{Abe:2016ero}, 
the total beam exposure is $27 \times 10^{21}$~p.o.t. and
the HK design proposes the construction of two identical 
water Cherenkov detectors in stage with fiducial mass of
187~kt per detector. The possibility of placing the first
detector near the Super-Kamiokande site, 295~km away
and $2.5\degree$ off-axis from the J-PARC neutrino beam
and the second detector in Korea having a baseline of 
1100~km from J-PARC at an off-axis angle of 
$\sim 2.5\degree$ has also been explored in 
Ref.~\cite{Abe:2016ero}, and this set-up has been referred 
as ``T2HKK". We follow the details as given in 
Ref.~\cite{Abe:2016ero} to simulate the T2HKK
set-up.

\subsection{Event Rates}
\label{subsec:event-rates}

In this subsection, we present the expected total event rates
for both the set-ups under consideration. We compute the 
number of expected electron events\footnote{The number of 
positron events can be calculated with the help of 
eq.~(\ref{eq:events}), by considering relevant oscillation 
probability and cross section. The same is valid for 
$\mu^{\pm}$ events.} in the $i$-th energy bin in the detector 
using the following well-known expression:
\begin{equation}
N_{i} = T\, n_n\, \epsilon\, \int_0^{E_{\rm max}}
dE \int_{E_{A_i}^{\rm min}}^{E_{A_i}^{\rm max}} dE_A \,\phi(E)
\,\sigma_\nue(E) \,R(E,E_A)\, P_{\mu e}(E) \, ,
\label{eq:events}
\end{equation}
where $\phi(E)$ is the neutrino flux spectrum 
($\mathrm{m}^{-2} \cdot \mathrm{year}^{-1} \cdot \mathrm{GeV}^{-1}$), 
$T$ is the total running time (year), 
$n_n$ is the number of target nucleons in the detector, $\epsilon$ 
is the detector efficiency, and $R(E,E_A)$ is the Gau\ss{}ian energy 
resolution function ($\mathrm{GeV}^{-1}$) of the detector. 
$\sigma_\nue$ is the neutrino 
interaction cross section ($\mathrm{m}^2$) which has been taken from 
Refs.~\cite{Messier:1999kj,Paschos:2001np}, where the authors 
calculated the cross section for water and isoscalar targets.
In order to have LAr cross sections for DUNE, we have scaled the 
inclusive charged current (CC) cross sections of water by a factor 
of 1.06 (0.94) for neutrino (antineutrino)~\cite{zeller,petti-zeller}.
We denote the true and reconstructed neutrino energies
by the quantities $E$ and $E_A$, respectively.

Table~\ref{tab:all-channels-signal-background-rates} shows a
comparison between the expected total signal and background 
event rates\footnote{While estimating these event rates, 
we properly consider the ``wrong-sign'' components which are 
present in the beam for both $\nue/\anue$ and $\numu/\anumu$ 
candidate events.} in the appearance and disappearance modes
for DUNE and T2HK set-ups. We compute the same for both 
neutrino and antineutrino runs assuming a total exposure of
248~$\mathrm{kt} \cdot \mathrm{MW} \cdot \mathrm{year}$ for DUNE and 
4200~$\mathrm{kt} \cdot \mathrm{MW} \cdot \mathrm{year}$ for T2HK. 
We consider a run-time ratio of 1:1 among neutrino and antineutrino
modes in DUNE and the corresponding ratio is 1:3 in T2HK. 
The total event rates are calculated assuming NO, 
$\dcp = 248\degree$, and $\sin^2\theta_{23}$ = 0.425.
For all other oscillation parameters, we consider the best fit values 
which are applicable for NO (see Table~\ref{tab:oscparams}).
To compute the full three-flavour neutrino oscillation probabilities in matter, 
we take the line-averaged constant Earth matter density of 2.80 (2.87) 
g/cm$^{3}$ for the T2HK (DUNE) baseline following the Preliminary
Reference Earth Model (PREM)~\cite{Dziewonski:1981xy}.

The main sources of backgrounds while selecting the $\nue$ and 
$\anue$ candidate events are the intrinsic $\nue$/$\anue$ component
in the beam, the muon events which will be mis-identified as 
electron events, and the neutral current (NC) events.
Table~\ref{tab:all-channels-signal-background-rates} clearly
depicts that in case of appea\-rance searches, the dominant 
background component is the intrinsic $\nue$/$\anue$ in the beam.
For the $\numu$/$\anumu$ candidate events, the main backgrounds
are the NC events. Though we present the total event rates in 
Table~\ref{tab:all-channels-signal-background-rates}, but, in our
simulation, we have performed a full spectral analysis using
the binned events spectra for both the DUNE and T2HK set-ups.

\begin{table}
\begin{center}
{
\newcommand{\mc}[3]{\multicolumn{#1}{#2}{#3}}
\newcommand{\mr}[3]{\multirow{#1}{#2}{#3}}
\begin{adjustbox}{width=1\textwidth}
\begin{tabular}{|c||c|c||c|c|}
\hline
\mr{3}{*}{\bf Mode (Channel)} & \mc{2}{c||}{\bf DUNE (248~$\bo{\mathrm{kt} \cdot \mathrm{MW} \cdot \mathrm{year}}$)} &\mc{2}{c|}{\bf T2HK (4200~$\bo{\mathrm{kt} \cdot \mathrm{MW} \cdot \mathrm{year}}$) } \\
\cline{2-5}
      & Signal & Background &Signal & Background \\
\cline{2-5}
      & CC & Int+Mis-id+NC=Total &CC &Int+Mis-id+NC=Total  \\
\hline \hline
$\nu$ (appearance) & 614 & 125+29+24=178 &2852 &530+13+173=716 \\
\hline
$\nu$ (disappearance) & 5040 & 0+0+24=24 &20024 &12+44+1003=1059 \\
\hline \hline
$\bar\nu$ (appearance) & 60 & 43+10+7=60 &1383 &627+11+265=903 \\
\hline
$\bar\nu$ (disappearance) & 1807 & 0+0+7=7 & 27447 &14+5+1287=1306 \\
\hline
\end{tabular}
\end{adjustbox}
}
\caption{Total signal and background event rates for DUNE and
T2HK set-ups assuming NO, $\dcp = 248\degree$, 
and $\sin^2\theta_{23}$ = 0.425. For all other oscillation parameters, 
we take the best fit values corresponding to NO (see Table~\ref{tab:oscparams}).
Here ``Int'' means intrinsic beam contamination, ``Mis-id'' represents 
mis-identified muon events, and ``NC'' stands for neutral current. 
See text for other details.}
\label{tab:all-channels-signal-background-rates}
\end{center}
\end{table}

\section{Details of Statistical Analysis}
\label{sec:analysis}

This section is devoted to describe the strategy that we adopt 
for the statistical treatment to quantify the sensitivities of DUNE
and T2HK in testing various lepton mixing schemes.
To produce our results, we take the help of the widely used 
GLoBES software~\cite{Huber:2004ka,Huber:2007ji}
which calculates the median sensitivity of the experiment without 
considering the statistical fluctuations. Unless stated otherwise,  
we generate our simulated data considering the best fit values 
of the oscillation parameters obtained in the global analysis assuming 
NO for the neutrino mass spectrum (see second column of 
Table~\ref{tab:oscparams}).
We also keep the choice of the neutrino mass ordering to be fixed 
to NO in the fit\footnote{DUNE will operate at multi-GeV energies
with 1300 km baseline and therefore, the matter effect is quite
substantial for this set-up. For this reason, DUNE can break the
hierarchy-$\dcp$ degeneracy 
completely~\cite{Agarwalla:2013hma} and can resolve the issue of
neutrino mass ordering at more than 5$\sigma$
C.L.~\cite{Acciarri:2015uup,Ballett:2016daj}.
Due to the shorter baseline of 295 km, T2HK has lower sensitivity
to the mass ordering. However, HK can settle this issue using the
atmospheric neutrinos at more than 3$\sigma$ C.L. for both NO 
and IO provided $\sin^2\theta_{23} > 0.45$~\cite{Yokoyama:2017mnt}.
Combining beam and atmospheric neutrinos in HK, the mass ordering 
can be determined at more than 3$\sigma$ (5$\sigma$) with five (ten)
years of data~\cite{Yokoyama:2017mnt}.}. The solar and atmospheric 
mass squared differences are already very well 
measured~\cite{Esteban:2016qun,Capozzi:2017ipn,deSalas:2017kay}
and moreover, they do not appear in the sum rule 
(see eq.~\eqref{eq:cosdelta}) that relates $\cos\dcp$
with the mixing angles. Therefore, we also keep them fixed in the fit
at their best fit values while showing our results.
Only in Appendix~\ref{app:dm31sq-marginalisation}, we give a
plot, where we marginalise over test $\Delta m_{31}^{2}$ in the fit
in its present 3$\sigma$ allowed range of $(2.45 - 2.69) \times 10^{-3}$
eV$^2$. The mixing angles play an important role in the 
sum rule and we treat them very carefully in our analysis. In the fit, 
we marginalise over test $\sin^2\theta_{23}$ in its present 3$\sigma$ 
allowed range of 0.381 to 0.615. We show few results where we also 
vary the true value of $\sin^2\theta_{23}$ or marginalise over it in the 
same 3$\sigma$ range. We do not impose any external prior on 
$\sin^2\theta_{23}$ as it will be directly measured by the DUNE and T2HK
experiments. The sum rule as given in eq.~\eqref{eq:cosdelta}
is very sensitive to the value of $\sin^2\theta_{12}$. We vary this
parameter in the fit in its present 3$\sigma$ allowed range of 
0.25 to 0.354. Since both DUNE and T2HK cannot constrain the 
solar mixing angle (see the probability expressions 
in~\cite{Akhmedov:2004ny}), we impose an external Gau\ss{}ian
prior of 0.7\% (at 1$\sigma$) on this parameter as the proposed
medium-baseline reactor oscillation experiment JUNO will be
able to measure $\sin^2\theta_{12}$ with this precision~\cite{An:2015jdp}.
We also marginalise over test $\sin^2\theta_{13}$ in its present 
3$\sigma$ allowed range of 0.019 to 0.024. While doing so, we 
apply an external Gau\ss{}ian prior of 3\% (at 1$\sigma$) on this 
parameter expecting that the Daya Bay experiment would be 
able to achieve this precision by the end of its run~\cite{Ling:2016wgq}.
Both DUNE and T2HK can measure $\theta_{13}$ with high precision
using $\numu \to \nue$ and $\anumu \to \anue$ oscillation channels 
and therefore, the prior on $\theta_{13}$ is not very crucial in our
study (see Fig.~\ref{fig:priors} and related discussion in 
Appendix~\ref{app:priors}). Still, we use this prior in our analysis
to speed up the marginalisation procedure. For a given test choice
of $\theta_{12}$, $\theta_{13}$, and $\theta_{23}$, the test value
of $\dcp$ is calculated using the sum rule
(see eq.~\eqref{eq:cosdelta}) for a particular choice of the lepton 
mixing scheme which is characterised by a certain value of 
$\theta^{\nu}_{12}$. Since the best fit value of $\dcp$
may change in the future, we also show some results varying the true
choice of $\dcp$ in the range 180$\degree$ to 
360$\degree$.

To perform our statistical analysis, we follow the procedure outlined 
in Refs.~\cite{Huber:2002mx,Fogli:2002au}, and use the following 
Poissonian $\chi^2$ function:
\begin{eqnarray}
\chi^2 = \min_{\xi_s, \xi_b}\left[2\sum^{n}_{i=1}
(\tilde{y}_{i}-x_{i} - x_{i} \ln \frac{\tilde{y}_{i}}{x_{i}}) +
\xi_s^2 + \xi_b^2\right ]\,,
\label{eq:chipull}
\end{eqnarray}
where $n$ is the total number of reconstructed energy bins and
\begin{eqnarray}
\tilde{y}_{i}(\{\omega\},\{\xi_s, \xi_b\}) = N^{th}_i(\{\omega\}) \left[
1+ \pi^s \xi_s \right] +
N^{b}_i(\{\omega\}) \left[1+ \pi^b \xi_b \right]\,.
\label{eq:rth}
\end{eqnarray}
Above, $N^{th}_i(\{\omega\})$ denotes the predicted number of CC 
signal events (estimated using eq.~\eqref{eq:events}) in the $i$-th 
energy bin for a set of oscillation parameters $\omega$. 
$N^{b}_i(\{\omega\})$ stands for the total number of background
events\footnote{Note that we consider both CC and NC background 
events in our analysis and the NC backgrounds are independent of
oscillation parameters.}. The quantities $\pi^s$ and $\pi^b$ in 
eq.~\eqref{eq:rth} represent the systematic errors on the signal 
and background events, respectively. For DUNE (T2HK), 
we consider $\pi^s$ = 5\% (5\%) and $\pi^b$ = 5\% (10\%)
in the form of normalisation errors for both the appearance 
and disappearance channels. We take the same uncorrelated
systematic uncertainties for both the neutrino and antineutrino 
modes. The quantities $\xi_s$ and $\xi_b$ denote the ``pulls'' 
due to the systematic error on the signal and background, 
respectively. The data in eq.~\eqref{eq:chipull} are included 
through the variable $x_i=N_i^{ex}+N_i^b$, where 
$N_i^{ex}$ is the number of observed CC signal events 
and $N_i^b$ is the background as discussed earlier.
To obtain the total $\chi^2$, we add the $\chi^2$
contributions coming from all the relevant oscillation 
channels for both neutrino and antineutrino modes
in a given experiment in the following fashion:
\begin{eqnarray}
\chi^2_{\rm total} &=&
\chi^2_{\numu \rightarrow \nue} + \chi^2_{\anumu \rightarrow \anue} + \chi^2_{\numu \rightarrow \numu} + \chi^2_{\anumu \rightarrow \anumu} + \chi^2_{\rm prior} \, .
\label{eq:totchisq}
\end{eqnarray}
In the above expression, we assume that all the oscillation channels for
both neutrino and antineutrino modes are completely uncorrelated, 
all the energy bins in a given channel are fully correlated, and the 
systematic errors on signal and background are fully uncorrelated.
The fact that the flux normalisation errors in $\nu_\mu \to \nu_e$ 
and $\nu_\mu \to \nu_\mu$ oscillation channels are same 
(i.e., they are correlated) is taken into account in the error budget 
for each of the two channels. However, there are other uncertainties
which contribute to the total normalisation error for each of the two 
channels, like the uncertainties in cross sections, detector 
efficiencies, etc., which are uncorrelated. For this reason, 
we simply assume that the total normalisation errors in these 
two channels are uncorrelated. The same is true for 
$\bar\nu_\mu \to \bar\nu_e$ and $\bar\nu_\mu \to \bar\nu_\mu$
oscillation channels. In our opinion, with the current understanding
of the two detectors in DUNE and T2HK experiments, it is premature 
to perform a very detailed analysis taking into account such fine effects 
as, e.g., the correlation between the flux normalisation errors in the 
appearance and disappearance channels.

In eq.~\eqref{eq:totchisq}, the last term appears due 
to the external Gau\ss{}ian priors that we impose on
$\sin^2\theta_{12}$ and $\sin^2\theta_{13}$ in the following way:
\begin{equation}
\chi^2_{\rm prior} = \left(\frac{
\sin^2\theta_{12} - \sin^2\theta_{12}^{\rm true}}
{\sigma(\sin^2\theta_{12})} \right)^2 +
\left(\frac{
\sin^2\theta_{13} - \sin^2\theta_{13}^{\rm true}}
{\sigma(\sin^2\theta_{13})} \right)^2 \, ,
\label{eq:prior}
\end{equation}
where we take $\sigma(\sin^2\theta_{12})$ = 0.007 $\times$ $\sin^2\theta_{12}^{\rm true}$
and $\sigma(\sin^2\theta_{13})$ = 0.03 $\times$ $\sin^2\theta_{13}^{\rm true}$
as mentioned earlier. In our analysis, we assume that 
the true values of $\sin^2\theta_{12}$ and $\sin^2\theta_{13}$
will remain unchanged in the future. While implementing the minimisation
procedure, $\chi^2_{\rm total}$ is first minimised with respect 
to the ``pull'' variables $\xi_s$, $\xi_b$, and then marginalised
over the various oscillation parameters in their allowed ranges 
in the fit as discussed above to obtain $\Delta\chi^2_{\textrm{min}}$.
In Fig.~\ref{fig:th23delta} in Appendix~\ref{app:ss23delta}, 
we quote the statistical significance 
of our results for 1 d.o.f. in terms of $n\sigma$, 
where $n=\sqrt{\Delta\chi^2_{\textrm{min}}}$, which is valid 
in the frequentist method of hypothesis testing~\cite{Blennow:2013oma}.

\section{Results and Discussion}
\label{sec:results}

\subsection{Compatibility between Various Symmetry Forms 
and Present Neutrino Oscillation Data}
\label{sec:compatibility}

\begin{figure}
\centering
\includegraphics[width=10cm]{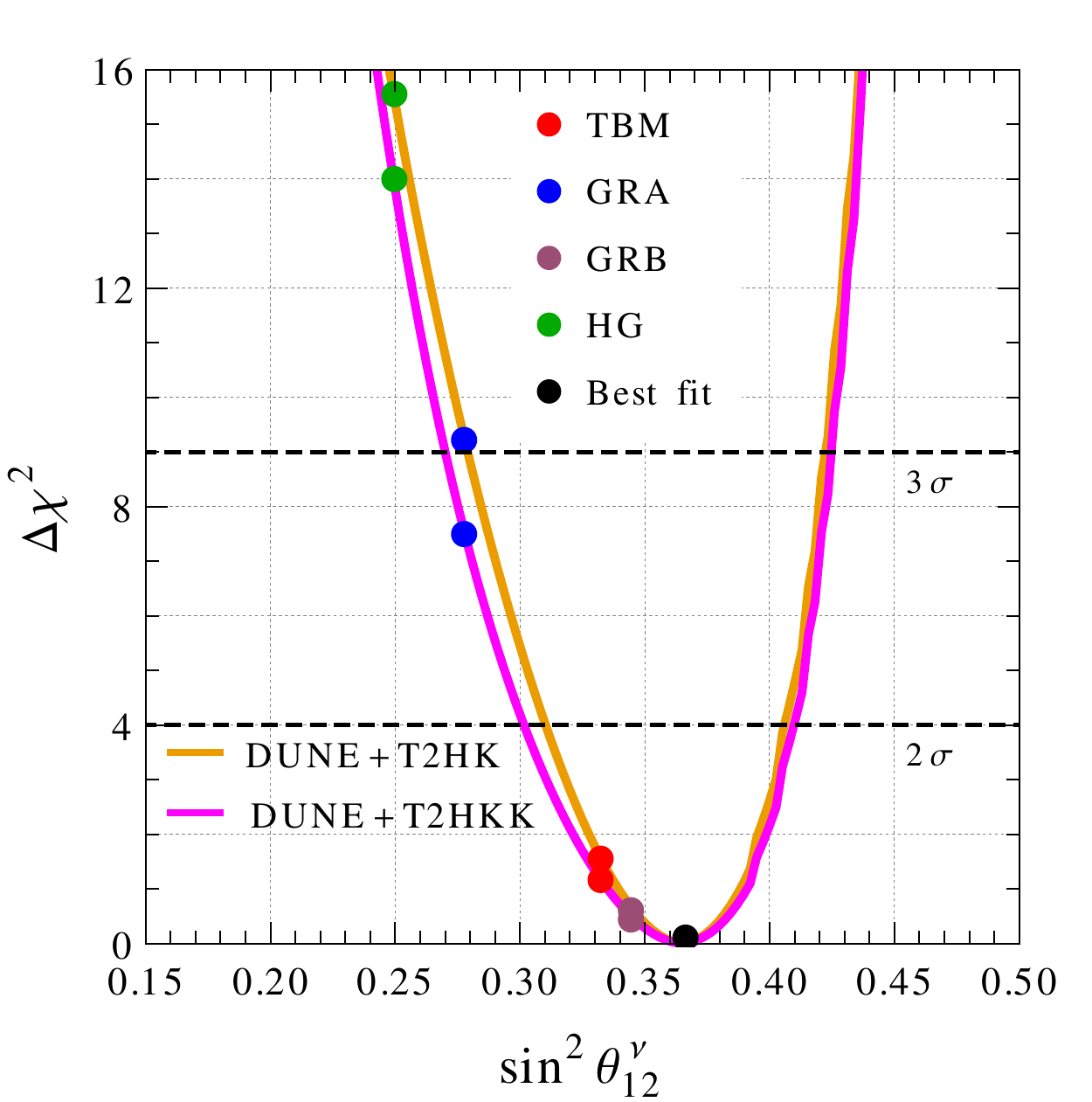}
\caption{Status of the lepton mixing symmetry forms 
in light of the current best fit values of the mixing angles and the CPV phase $\dcp$ 
and using the potential of DUNE\,+\,T2HK and DUNE\,+\,T2HKK.
The black dot corresponds to $\sin^2\th^\nu_{12} = 0.364$ obtained from 
eq.~\eqref{eq:cosdelta} using the present best fit values of 
$\th_{12}$, $\th_{13}$, $\th_{23}$, and $\dcp$. 
The coloured dots correspond to the values of $\sin^2\th^\nu_{12}$ for the 
TBM, GRA, GRB, and HG symmetry forms.}
\label{fig:currentBFV}
\end{figure}

Table~\ref{tab:oscparams} shows the current best fit values of 
the neutrino mixing angles and the CPV phase $\dcp$. 
Equation~\eqref{eq:cosdelta} relates 
$\dcp$ with the parameter $\th^\nu_{12}$ which characterises  the symmetry forms 
under consideration. 
The first question we want to answer is how much compatibility there is between the 
mixing symmetry forms and the present best fit values of the oscillation parameters.  
To this aim, we assume the current best fit values 
of $\th_{12}$, $\th_{13}$, $\th_{23}$, and $\dcp$ to be their true values 
and generate prospective data using the DUNE, T2HK, and T2HKK
experimental set-ups according to the procedure explained in Section~\ref{sec:analysis}. 
Further, in the test, we fix the mixing angles to their best fit values and let 
$\dcp$ to vary between $180\degree$ and $360\degree$. 
For each value of $\dcp$, first we calculate $\sin^2\th^\nu_{12}$ according to 
eq.~\eqref{eq:cosdelta}, and then, we estimate the corresponding $\Delta\chi^2$ 
using the prospective data from combined DUNE\,+\,T2HK 
and DUNE\,+\,T2HKK set-ups.  
This is $\Delta\chi^2$ between a given value of $\dcp$ and 
its best fit value, i.e., $\dcp = 248\degree$. 
In Fig.~\ref{fig:currentBFV}, we plot $\Delta\chi^2$ as a function of 
$\sin^2\th^\nu_{12}$ and show on the resulting curves:
\begin{itemize}
\item the black dot corresponding to the current best fit value of $\dcp = 248\degree$ 
which translates to $\sin^2\th^\nu_{12} = 0.364$ ($\Delta\chi^2 = 0$); 
\item the coloured dots corresponding to the values of $\sin^2\th^\nu_{12}$ 
which characterise the GRB (violet), TBM (red), GRA (blue) and HG (green) symmetry forms. 
\end{itemize}
From this figure we see that, if the present best fit values 
of $\th_{12}$, $\th_{13}$, $\th_{23}$, and $\dcp$ 
were the true values of these parameters, 
the GRB (TBM) symmetry form would be compatible with 
them at slightly less (more) than $1\sigma$ C.L.,
while the GRA and HG schemes would be disfavoured
at more than $2.7\sigma$ and $3.7\sigma$, respectively, 
for both the combined set-ups.

However, at present the CPV phase $\dcp$ is not severely constrained, and 
as can be seen from Table~\ref{tab:oscparams}, any value between 
$180\degree$ and $342\degree$ is allowed at $2\sigma$ C.L., 
and any value except for the ones between $31\degree$ and $137\degree$ 
is allowed at $3\sigma$. 
Fixing the three mixing angles to their best fit values, 
we find from eq.~\eqref{eq:cosdelta} that the full range of $\cos\dcp \in [-1,1]$ 
(allowed at present at $3\sigma$) translates to the values of 
$\sin^2\th^\nu_{12} \in [0.157, 0.460]$. 
Thus, in principle, any value from this range may turn out to be favoured in the future.
For instance, imagine that in the future the best fit value of $\dcp$ will shift from 
$248\degree$ to $290\degree$, while the best fit values of the mixing angles will 
remain the same. Then, the value of $\sin^2\th^\nu_{12} = 0.250$, 
and thus the HG symmetry form, will be favoured. 
With this said, one should keep in mind that the position of the black dot in 
Fig.~\ref{fig:currentBFV} is likely to change in the future, 
but having more precise measurements of $\dcp$ and the mixing angles at our disposal, 
we will be able to repeat this analysis 
favouring some symmetry forms and disfavouring the others. 

\begin{figure}
\centering
\includegraphics[width=10cm]{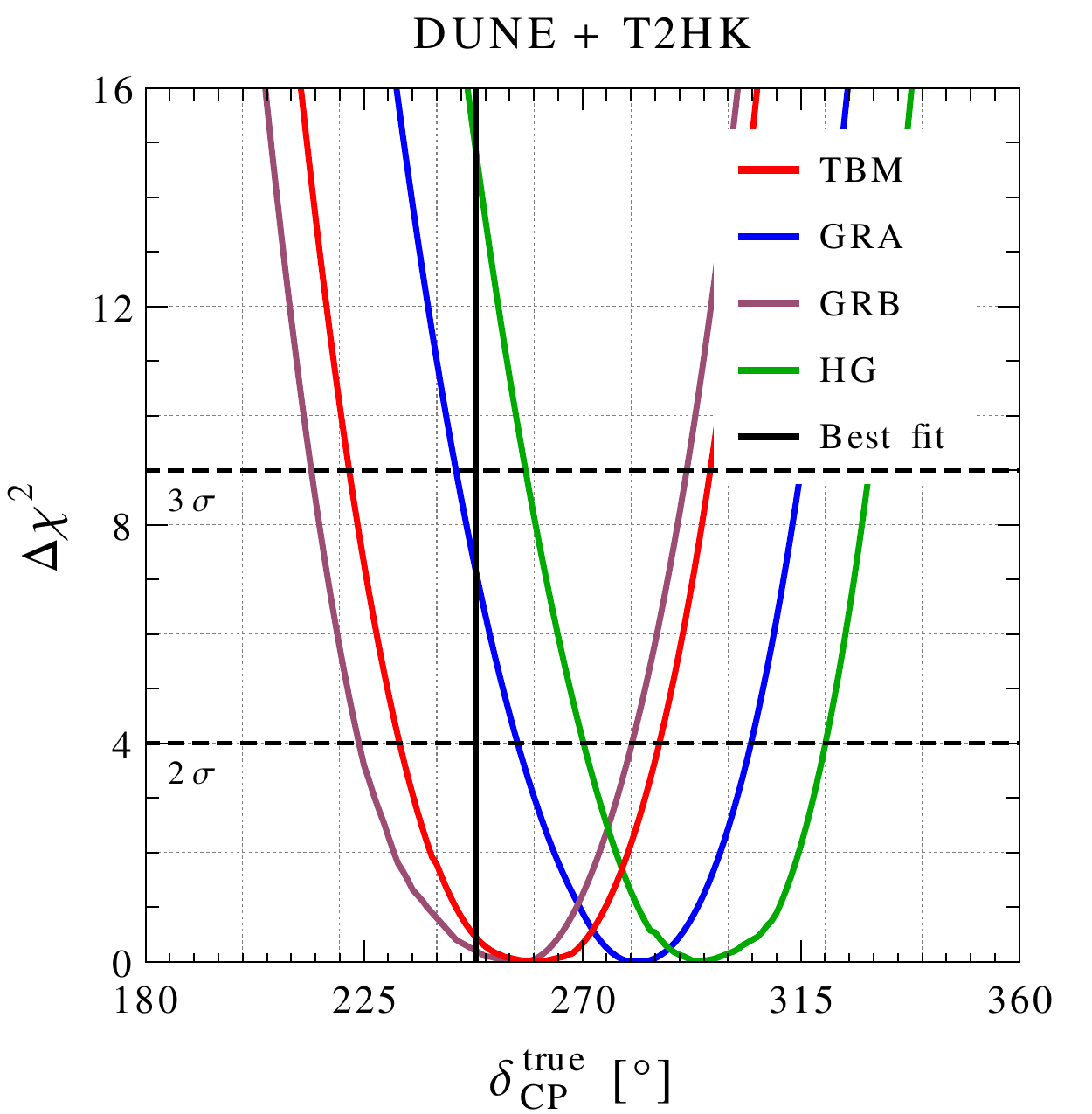}
\caption{Compatibility of the TBM, GRA, GRB, and HG symmetry forms 
with any potentially true value of the Dirac CPV phase $\dcp$. 
The figure is obtained employing combined potential of DUNE and T2HK.
The black vertical line corresponds to the current best fit value of $\dcp$ for 
the NO neutrino mass spectrum.}
\label{fig:chi2vsdelta}
\end{figure}
 Having obtained an idea of how much the mixing symmetry forms in question are 
compatible with the present best fit values of the oscillation parameters, 
we go next to a more involved analysis which will allow us to see the compatibility 
of the studied symmetry forms with any value of $\dcp$ between 
$180\degree$ and $360\degree$, should it turn out to be the true value. 
To this aim, we fix the true value of the CPV phase, $\dcp^{\rm true}$, to be 
between $180\degree$ and $360\degree$, 
the true value of the atmospheric mixing angle, $\th_{23}^{\rm true}$,
to a value from its $3\sigma$ range, 
and the true values of the solar and reactor mixing angles, 
$\th_{12}^{\rm true}$ and $\th_{13}^{\rm true}$, to their corresponding best fit values. 
Then, we generate data with this input 
using the DUNE and T2HK set-ups. 
In the test, we assume a given symmetry form to hold and fix the three test values 
$\th_{12}^{\rm test}$, $\th_{13}^{\rm test}$, and $\th_{23}^{\rm test}$ to values from the corresponding $3\sigma$ ranges. 
Using these test values and known for a given symmetry form $\th^\nu_{12}$, 
we calculate $\dcp^{\rm test}$ from eq.~\eqref{eq:cosdelta}.
Each couple of the true and test oscillation vectors, 
$\bo{\rm y^{true}} = (\th_{12}^{\rm true}, \th_{13}^{\rm true}, \th_{23}^{\rm true}, \dcp^{\rm true})$ and 
$\bo{\rm y^{test}} = (\th_{12}^{\rm test}, \th_{13}^{\rm test}, \th_{23}^{\rm test}, \dcp^{\rm test})$, provides a certain value of $\Delta\chi^2$.
Note that in calculating this value, we use external priors on 
$\sin^2\th_{12}$ and $\sin^2\th_{13}$ 
from JUNO and Daya Bay, as explained in Section~\ref{sec:analysis}.
A detailed discussion on the impact of these two priors 
is presented in Appendix~\ref{app:priors}. 
Further, for each $\bo{\rm y^{true}}$ we marginalise 
over $\th_{ij}^{\rm test}$ (over $\bo{\rm y^{test}}$). 
Finally, for each $\dcp^{\rm true}$ we marginalise also over $\th_{23}^{\rm true}$. 
We repeat this procedure for each of the four symmetry forms in study. 
The obtained results are shown in Fig.~\ref{fig:chi2vsdelta}.
Two comments on Fig.~\ref{fig:chi2vsdelta} are in order.
\begin{itemize}
\item 
For each symmetry form a significant part of the parameter space gets disfavoured 
at more than $3\sigma$. Should the true value of $\dcp$ lie in this part of 
the parameter space, the corresponding symmetry form will be disfavoured 
at $3\sigma$ confidence level.
\item 
Now we can see at which C.L. any given symmetry form is compatible with any 
potentially true value of $\dcp$. We just need to draw a vertical line at $\dcp^{\rm true}$ 
of interest. The points where it crosses the $\Delta\chi^2$ curves will provide the 
confidence levels at which the TBM, GRA, GRB, and HG forms are compatible with 
this $\dcp^{\rm true}$. In particular, for $\dcp^{\rm true} = 248\degree$, we find numbers 
which correspond to those extracted from 
Fig.~\ref{fig:currentBFV}\footnote{Note that the numbers we read from 
Fig.~\ref{fig:chi2vsdelta} are slightly smaller than those extracted from 
Fig.~\ref{fig:currentBFV} due to the fact that now we marginalise over 
the values of the mixing angles.}.
\end{itemize}

\subsection{How well can DUNE and T2HK Separate between Various Symmetry Forms?}
\label{sec:distinguishing}

 In this subsection, we will answer the question of how well DUNE and T2HK 
can distinguish the discussed symmetry forms under the assumption that one of them 
is realised in Nature. 
Given the fact that the BM form is not compatible with the current best fit values of 
the neutrino mixing angles, which we are going to use first in our analysis, 
we end up with four best fit values of interest. 
Namely, from Table~\ref{tab:deltaBF}, we read $\dcp = 256\degree$, 
$261\degree$, $282\degree$, and $293\degree$ for the GRB, TBM, GRA, and HG 
symmetry forms, respectively. 
Assuming one of them to be the true value of $\dcp$, we will test the 
remaining three values against the assumed true value 
using DUNE, T2HK, and their combination. 
Overall, we have 12 pairs of the values we want to compare.

We start with DUNE. After performing a statistical analysis of simulated data, 
as described in Section~\ref{sec:analysis}, 
we obtain that for all the 12 cases $\D\chi^2$ does not exceed approximately $3.5$.
This value of $\D\chi^2$ is found when the value of $\dcp$ predicted in the HG (GRB) 
case is tested against the value of $\dcp$ for the GRB (HG) form, which 
is assumed to be the true one. 
Therefore, the sensitivity of DUNE alone is not enough to make a $3\sigma$ claim on 
discriminating between the symmetry mixing forms under investigation, 
and we will test next all the cases using simulated data from the T2HK experiment, 
whose overall sensitivity to CPV is better than that of DUNE.

Performing a statistical analysis for T2HK, we find that it can discriminate 
the GRB case from the HG case at approximately $2.5\sigma$ confidence level. 
More specifically, 
if $\dcp = 256\degree~(293\degree)$ turned out to be 
the true value of the CPV phase, then T2HK could disfavour 
the value of $\dcp = 293\degree~(256\degree)$ with $\D\chi^2 \approx 7.5$.
We also find that the TBM and HG symmetry forms, in turn, 
occur to be resolvable at slightly less C.L.  
with $\D\chi^2$ being around $5.5$.
Thus, the sensitivity of T2HK is not sufficient as well to discriminate between 
the cases of interest at $3\sigma$ C.L. 
For that reason, we will test them further using the potential of 
combining DUNE and T2HK.

The combination of DUNE and T2HK provides better sensitivity to the CPV phase $\dcp$ 
than either of these two experiments in isolation (see, e.g., \cite{Ballett:2016daj}).
A combined analysis performed by us leads to the results described below.
Firstly, the GRB and HG mixing forms can be now distinguished 
at more than $3\sigma$ confidence level.
If $\dcp = 256\degree~(293\degree)$ is the true value, 
then $\dcp = 293\degree~(256\degree)$ will be disfavoured with 
$\D\chi^2 \approx 11~(10.5)$.
Secondly, the TBM and HG cases can be resolved at slightly less than $3\sigma$, 
the corresponding values of $\D\chi^2$ being around $8$. 
Thirdly, discriminating between the GRA and GRB forms can be claimed 
with $\D\chi^2 \approx 5.5$. 
Finally, the sensitivity of the combination of these two experiments is not enough 
to discern TBM from GRA, GRA from HG, and TBM from GRB at even $2\sigma$. 
For these three pairs, we find $\D\chi^2 \approx 3.5$, $1.2$, and $0.2$, respectively, 
when the corresponding predictions for $\dcp$ are compared between themselves.

We have checked that adding NO$\nu$A and T2K data to sets of simulated data 
obtained using the DUNE and T2HK set-ups 
leads to increase in the values of $\D\chi^2$ of only several tenths. 
Thus, inclusion of these data does not help to improve differentiating between 
the considered mixing schemes. 
We summarise the obtained results in Table~\ref{tab:summary}, in which
we present confidence levels (in number of $\sigma$) at which the symmetry forms 
under consideration can be distinguished from each other 
assuming that one of them is realised in Nature and using the potential of 
combining DUNE with T2HK. 
\begin{table}
\centering
\renewcommand{\arraystretch}{1.2}
\begin{tabular}{|l|llll|}
\hline
\diagbox{True}{Tested} & TBM & GRA & GRB & HG \\
\hline
\bt
TBM &  & $1.9$ & $0.5$ & $\bo{2.9}$ \\
GRA & $1.9$ &  & $2.3$ & $1.1$ \\
GRB & $0.5$ & $2.3$ &  & $\bo{3.3}$ \\
HG & $\bo{2.9}$ & $1.1$ & $\bo{3.3}$ & \\
\hline
\end{tabular}
\caption{Confidence levels (in number of $\sigma$) at which the symmetry forms 
under consideration can be distinguished from each other 
assuming that one of them is realised in Nature. 
The result is obtained using the combination DUNE\,+\,T2HK. 
All the mixing angles have been fixed to their NO best fit values both in data and in test.}
\label{tab:summary}
\end{table}

\begin{figure}[t!]
\centering
\includegraphics[width=\textwidth]{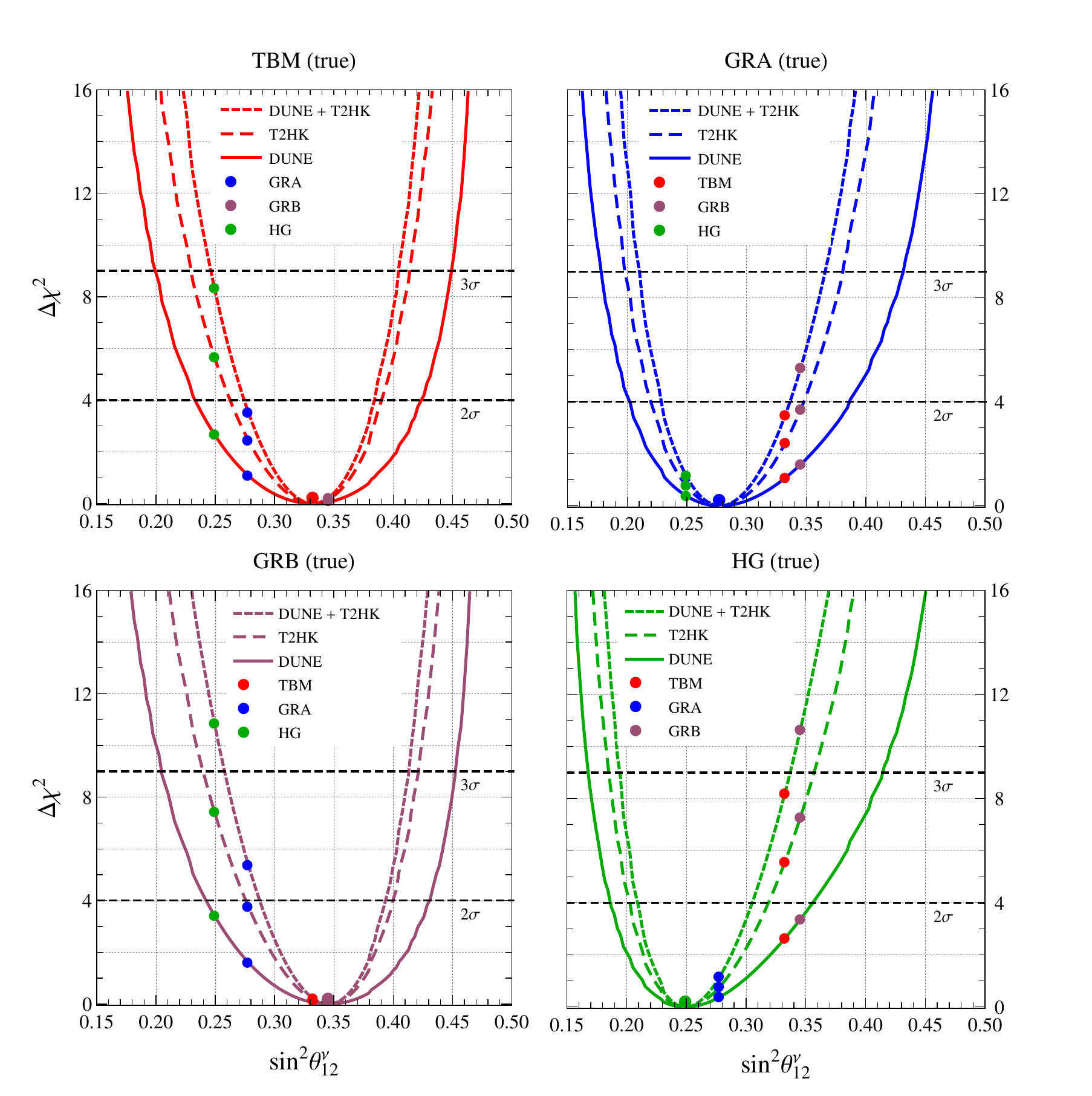}
\caption{Sensitivities of DUNE, T2HK, and their combination to distinguish between 
the TBM, GRA, GRB, and HG symmetry forms under the assumption that one of them 
is realised in Nature. In the top left (right) panel the assumed true symmetry form is 
TBM (GRA), while in the bottom left (right) panel this form is GRB (HG). 
The GRB (HG) form can be discriminated from HG (GRB) at more than $3\sigma$ C.L. 
using the combined potential of DUNE and T2HK.}
\label{fig:thnu12}
\end{figure}

Further, performing the more involved analysis described in Appendix~\ref{app:priors}, 
we obtain the results summarised in Fig.~\ref{fig:thnu12}. 
This figure allows us to see immediately at which C.L. a given pair of the symmetry forms 
can be distinguished, under the assumption that one form in the pair is realised in Nature. 
In particular, the numbers presented in Table~\ref{tab:summary} get clear 
graphic representation. Indeed, we see that using the combination DUNE\,+\,T2HK, 
GRB and HG can be resolved at more than $3\sigma$ C.L., 
while TBM and HG can be distinguished at almost $3\sigma$. 

 As we see from Appendix~\ref{app:priors}, the external prior on $\sin^2\th_{12}$ from 
JUNO is very important for the analyses performed in the present study. 
Usually, the present precision on $\sin^2\th_{12}$ is sufficient for the LBL experiments 
to achieve their goals on determination of $\dcp$, neutrino mass ordering, 
and the octant of $\th_{23}$. However, in our case, the role of $\th_{12}$ 
is very important, since, as we have mentioned earlier, eq.~\eqref{eq:cosdelta}, 
and thus predictions for $\dcp$ provided by different symmetry forms, 
are very sensitive to the value of the solar angle. 
Thereby, there is a nice synergy between JUNO on the one hand and 
the LBL experiments on the other: DUNE and T2HK will be much more sensitive 
in addressing the questions posed in the present study, 
if they are provided with a precise measurement of $\th_{12}$ performed by JUNO.

 Finally, we would like to notice that the $\Delta\chi^2$ values obtained in the case of 
DUNE\,+\,T2HK in Fig.~\ref{fig:thnu12} can also be inferred from Fig.~\ref{fig:chi2vsdelta}. 
Namely, drawing a vertical line at the minimum of $\Delta\chi^2$ curve for a given 
symmetry form in Fig.~\ref{fig:chi2vsdelta}, 
we can assess how much the other forms are disfavoured 
with respect to the chosen form. For example, let us assume that the HG form 
is realised in Nature. Then, we have $\dcp^{\rm true} = 293\degree$ 
(see Table~\ref{tab:deltaBF}). Drawing a vertical line at this value of $\dcp^{\rm true}$, 
we read from the intersections with the GRA, TBM, and GRB curves: 
$\Delta\chi^2 \approx 1$, $7$, and $10$, respectively. 
These are to be compared with the bottom right panel of Fig.~\ref{fig:thnu12}.

\subsection{The BM Symmetry Form}
\label{sec:BM}

 As we have mentioned in the Introduction, even though the BM symmetry form 
is not compatible with the current best fit values of the neutrino mixing angles, 
it turns out to be viable, if the current $2\sigma$ ranges of the mixing angles are 
taken into account. For example, if we keep $\sin^2\th_{13}$ and $\sin^2\th_{23}$ fixed 
to their best fit values for NO, we find that the value of $\sin^2\th_{12} = 0.3343$, 
which is the upper bound of the corresponding $2\sigma$ range 
(see Table~\ref{tab:oscparams}),
is required to obtain $\cos\dcp = -1.00$ and thus, recover viability of the BM mixing form.
For this choice of values of the mixing angles, the values of $\cos\dcp$ (and $\dcp$),
predicted in the TBM, GRA, GRB, and HG cases, change. 
We summarise them in Table~\ref{tab:BMviable}. 
\begin{table}
\centering
\renewcommand{\arraystretch}{1.2}
\begin{tabular}{|l|cc|}
\hline\btb
Symmetry form & $\cos\dcp$ & $\dcp$ \\
\hline\bt
BM & $-1.00$ & 180\degree \\
TBM & $\phantom{-}0.07$ & $86\degree \lor 274\degree$ \\
GRA & $\phantom{-}0.43$ & $65\degree \lor 295\degree$ \\
GRB & $-0.01$ & $91\degree \lor 269\degree$ \\
HG & $\phantom{-}0.60$ & $53\degree \lor 307\degree$ \\
\hline
\end{tabular}
\caption{The values of $\cos\dcp$ and $\dcp$ 
for different symmetry forms obtained from the sum rule in eq.~\eqref{eq:cosdelta} 
fixing $\sin^2\th_{12} = 0.3343$ (its upper $2\sigma$ bound) and 
two other mixing angles to their NO best fit values.}
\label{tab:BMviable}
\end{table}

We perform the analysis in this case and find that the BM form 
can be distinguished from all the other forms at more than $5\sigma$ 
by DUNE alone. The corresponding $\D\chi^2$ are between $25$ and $31$, 
and they translate to the numbers of $\sigma$ presented 
in Table~\ref{tab:summaryBM}. 
T2HK provides even better results, which we also show in 
Table~\ref{tab:summaryBM}. 
\begin{table}
\centering
\renewcommand{\arraystretch}{1.1}
\setlength\bigstrutjot{1.3ex}
\begin{tabular}{|l|lllll|}
\hline
\diagbox{True}{Tested} & BM & TBM & GRA & GRB & HG \\
\hline
\bt
\multirow{2}{*}{BM} &  & $\bo{5.1}$ (D) & $\bo{5.3}$ (D) & $\bo{5.0}$ (D) & $\bo{5.4}$ (D) \\ \bb
& & $\bo{9.4}$ (T) & $\bo{9.8}$ (T) & $\bo{9.2}$ (T) & $\bo{9.7}$ (T) \\ 
\bt
\multirow{2}{*}{TBM} & $\bo{5.2}$ (D) &   & 
   \multirow{2}{*}{$2.1$} & 
   \multirow{2}{*}{$0.5$} & 
   \multirow{2}{*}{$\bo{3.4}$} \\ \bb
& $\bo{8.9}$ (T) & & & & \\ 
\bt
\multirow{2}{*}{GRA} & $\bo{5.5}$ (D) & 
   \multirow{2}{*}{$2.1$} &  & 
   \multirow{2}{*}{$2.5$} & 
   \multirow{2}{*}{$1.4$} \\ \bb
& $\bo{9.2}$ (T) & & & & \\
\bt
\multirow{2}{*}{GRB} & $\bo{5.1}$ (D) & 
   \multirow{2}{*}{$0.5$} & 
   \multirow{2}{*}{$2.5$} &  & 
   \multirow{2}{*}{$\bo{3.1}$ (T)} \\ \bb 
& $\bo{8.7}$ (T) & & & & \\ 
\bt
\multirow{2}{*}{HG} & $\bo{5.6}$ (D) & 
   \multirow{2}{*}{$\bo{3.4}$} & 
   \multirow{2}{*}{$1.4$} & 
   \multirow{2}{*}{$\bo{3.1}$ (T)} & \\
& $\bo{9.2}$ (T) & & & & \\
\hline
\end{tabular}
\caption{Confidence levels (in number of $\sigma$) at which the symmetry forms 
under consideration can be distinguished from each other by different experiments 
in the case of possibility to have viable BM mixing in the neutrino sector. 
``D'' and ``T'' stand for DUNE and T2HK, respectively. 
When not explicitly specified, the results are for DUNE\,+\,T2HK. 
Both in data and in test, $\sin^2\th_{12}$ has been set to $0.3343$, 
while $\sin^2\th_{23}$ and $\sin^2\th_{13}$ have been fixed to their NO best fit values.}
\label{tab:summaryBM}
\end{table}

\section{Summary and Conclusions}
\label{sec:conclusions}

 In the present study, we have explored in detail 
the sensitivity of the future LBL experiments DUNE and T2HK to test 
various lepton mixing schemes predicted by flavour models 
with non-Abelian discrete symmetries. These models provide a 
natural explanation of the observed pattern of neutrino mixing.  
We have concentrated on a particular sum rule for 
$\cos\dcp$  given in eq. (\ref{eq:cosdelta}),
which holds for a rather broad class of discrete flavour symmetry models. 
We have considered five different underlying symmetry forms of the 
neutrino mixing matrix, namely,  
bimaximal (BM), tri-bimaximal (TBM), golden ratio type A (GRA), 
golden ratio type B (GRB), and hexagonal (HG).
Each of these mixing schemes is characterised by a specific value of the 
angle $\th^\nu_{12}$ entering into the sum rule in eq.~\eqref{eq:cosdelta}. 
The values of $\th^\nu_{12}$ for the BM, TBM, GRA, GRB, and HG forms 
are $45\degree$, $35\degree$, $32\degree$, $36\degree$, and $30\degree$, respectively. 
The BM symmetry form is disfavoured by the present best fit values of the mixing angles. 
Table~\ref{tab:deltaBF} summarises the predictions for
$\dcp$ for the other symmetry forms 
assuming the current best fit values of $\th_{12}$, $\th_{23}$, and $\th_{13}$.
In our analysis, we have considered only the predicted values of $\dcp$ 
lying around $270\degree$, 
since they are preferred by the present oscillation data (see Table~\ref{tab:oscparams}).

 Based on the prospective DUNE\,+\,T2HK data, 
the GRB and TBM symmetry forms are compatible with
the current best fit values of the mixing parameters at around $1\sigma$ confidence level. 
Under the same condition, the GRA and HG forms are disfavoured at around 
$3\sigma$ and $4\sigma$, respectively (see Fig.~\ref{fig:currentBFV}). 
Next, in Fig.~\ref{fig:chi2vsdelta}, we show  
up to what extent any given symmetry form is compatible with any true value of $\dcp$ 
lying in the range $180\degree$ to $360\degree$. 
In our analysis, we impose an external Gau\ss{}ian prior of $0.7\%$ 
(at $1\sigma$) on $\sin^2\th_{12}$ as expected from the upcoming
JUNO experiment, which improves our results significantly, 
as shown in Fig.~\ref{fig:priors} in Appendix~\ref{app:priors}.
This demonstrates a very important synergy 
between JUNO and LBL experiments like DUNE and T2HK, 
while testing various lepton mixing schemes in light of oscillation data.

 The combined data from DUNE and T2HK 
can discriminate among GRB and HG at more than $3\sigma$, 
if one of them is realised in Nature and the other form is
tested against it (see Table~\ref{tab:summary}). 
The same is true for TBM and HG at almost $3\sigma$.
Note, in these two cases, the differences between the predicted 
best fit values of $\dcp$ are $37\degree$ and $32\degree$, respectively 
(see Table~\ref{tab:deltaBF}).
Similarly, the GRA symmetry form can be distinguished from GRB and TBM 
at around $2\sigma$. The corresponding differences in these cases are 
$26\degree$ and $21\degree$, respectively.
At the same time, there is a difference of $11\degree$ 
for GRA and HG, which can be discriminated only at $1\sigma$.
For TBM and GRB, the difference is only $5\degree$ and therefore,
the significance of separation is very marginal (around $0.5\sigma$).
 
  In conclusion, the detailed analyses performed in the present work can be applied to 
any flavour model leading to a sum rule which predicts $\dcp$. 
In this regard, our article can serve as a useful guidebook for further studies.

\section*{Acknowledgements}

 S.K.A. and S.S.C. are supported by the DST/INSPIRE Research Grant [IFA-PH-12],
Department of Science \& Technology, India. A part of S.K.A.'s work was
carried out at the International Centre for Theoretical Physics (ICTP),
Trieste, Italy. It is a pleasure for him to thank the ICTP for the hospitality
and support during his visit via SIMONS Associateship.
A.V.T. and S.T.P. acknowledge funding from the European Union's Horizon 2020 
research and innovation programme
under the  Marie Sk\l{}odowska-Curie grant 
agreements No 674896 (ITN Elusives) and No 690575 (RISE InvisiblesPlus). 
This work was supported in part 
by the INFN program on Theoretical Astroparticle Physics (TASP) 
and by the  World Premier International Research Center 
Initiative (WPI Initiative, MEXT), Japan (S.T.P.).

\appendix

\section{Issue of Priors on $\bo{\sin^2\th_{12}}$ and $\bo{\sin^2\th_{13}}$}
\label{app:priors}

In this appendix, we discuss the role of external priors on various mixing angles that we considered in our analysis. 
First of all, we do not consider any prior on $\sin^2\th_{23}$ 
since both DUNE and T2HK will be able to measure this parameter with sufficient precision.
However, since these experiments are not sensitive to $\th_{12}$ 
(see probability expressions in~\cite{Akhmedov:2004ny}), we consider an external Gau\ss{}ian
prior of 0.7\% (at 1$\sigma$) on $\sin^2\theta_{12}$ as expected from the proposed
JUNO experiment~\cite{An:2015jdp}.
Even though both DUNE and T2HK can provide high precision measurement of $\theta_{13}$ 
using their appearance channels, but to speed up our computation, we also apply 
a Gau\ss{}ian prior of 3\% (at 1$\sigma$) on $\sin^2\th_{13}$ as expected by the end of Daya Bay's run~\cite{Ling:2016wgq}.

\begin{figure}[t]
\centering
\includegraphics[width=10cm]{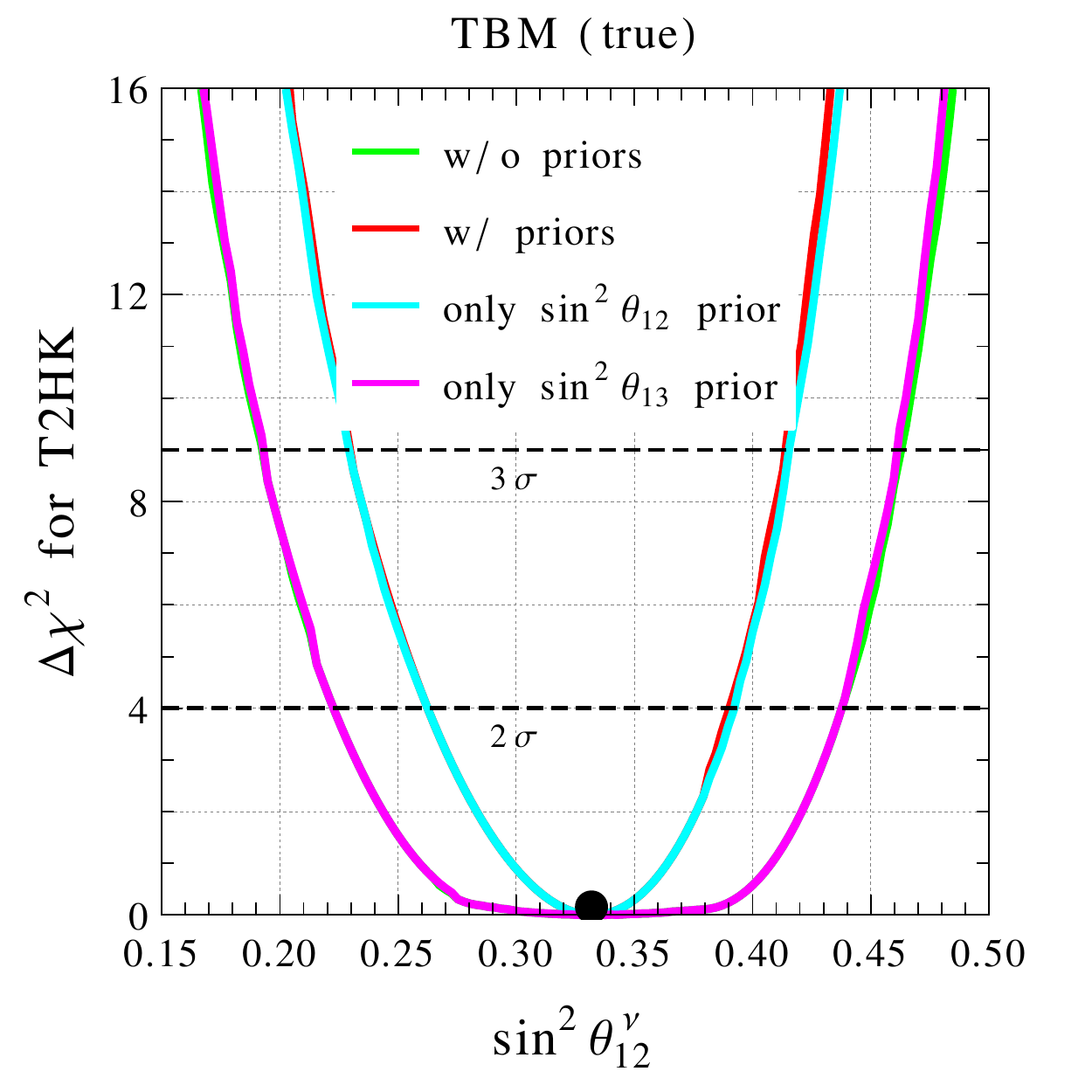}
\caption{Impact of external Gau\ss{}ian priors on $\sin^2\th_{12}$ and $\sin^2\th_{13}$ 
on the resulting $\Delta\chi^2$ function in the case of T2HK. 
We use a prior on $\sin^2\th_{12}$ of 0.7\% (at $1\sigma$) from JUNO 
and a prior on $\sin^2\th_{13}$ of $3\%$ (at $1\sigma$) from Daya Bay. 
While doing so, we assume the current best fit values of these two parameters to be 
their true values. 
The black dot corresponds to $\sin^2\th^\nu_{12} = 1/3$, 
which characterises the TBM symmetry form.}
\label{fig:priors}
\end{figure}

Fig.~\ref{fig:priors} shows the impact of these priors in our analysis. 
We obtain this figure in the following way.
First, we assume one of the symmetry forms, e.g., TBM,  
to be realised in Nature. For this symmetry form, we estimate the true value of $\dcp$ from eq.~\eqref{eq:cosdelta} using the current best fit values 
of the mixing angles for NO (see Table~\ref{tab:oscparams}) as their true values.
We generate data with this input. 
Then, in the test, we vary $\sin^2\th_{ij}$ and $\dcp$ in their corresponding $3\sigma$ allowed ranges. 
For each set of test values of $\sin^2\th_{ij}$ and $\dcp$, 
we estimate $\Delta\chi^2$ and also calculate 
the corresponding value of $\sin^2\th^\nu_{12}$ using eq.~\eqref{eq:cosdelta}.
For the same $\sin^2\th^\nu_{12}$, there can be several values of $\Delta\chi^2$. 
From there, for each $\sin^2\th^\nu_{12}$, 
we choose the minimal value of $\Delta\chi^2$. 
Finally, we plot this minimum $\Delta\chi^2$ as a function of $\sin^2\th^\nu_{12}$ in Fig.~\ref{fig:priors}. 
We present the results for T2HK considering the following cases:
(i) without assuming any priors on $\sin^2\th_{12}$ and $\sin^2\th_{13}$, 
(ii) assuming priors on both the parameters, 
(iii) only the prior on $\sin^2\th_{12}$, 
and (iv) only the prior on $\sin^2\th_{13}$.
From this figure, we can make the following few important observations. 
\begin{itemize}
\item First, we see that the curves corresponding to cases (i) and (iv) almost overlap 
with each other. It suggests that for the physics case under study, T2HK does not need an external prior on $\sin^2\th_{13}$ since it can provide a necessary precision on this parameter.
\item Secondly, we observe that the curves corresponding to cases (ii) and (iii) also overlap 
with each other. It indicates that for our purpose, T2HK needs an external prior on $\sin^2\th_{12}$ from JUNO since it has a very mild sensitivity on this mixing parameter.
\end{itemize}

We have checked that the above observations are also valid for GRA, GRB, 
and HG symmetry forms and for DUNE as well. We have also seen that 
cases (ii) and (iii) are almost equivalent to the fixed parameter scenario, 
where we keep all the mixing angles to be fixed at their best fit values 
in the fit. Unless mentioned otherwise, we always impose both these 
priors in our statistical analysis, as described in Section~\ref{sec:analysis}.

\section{Impact of Marginalisation over $\bo{\Delta m_{31}^{2}}$}
\label{app:dm31sq-marginalisation}

\begin{figure}[t]
\centering
\includegraphics[width=10cm]{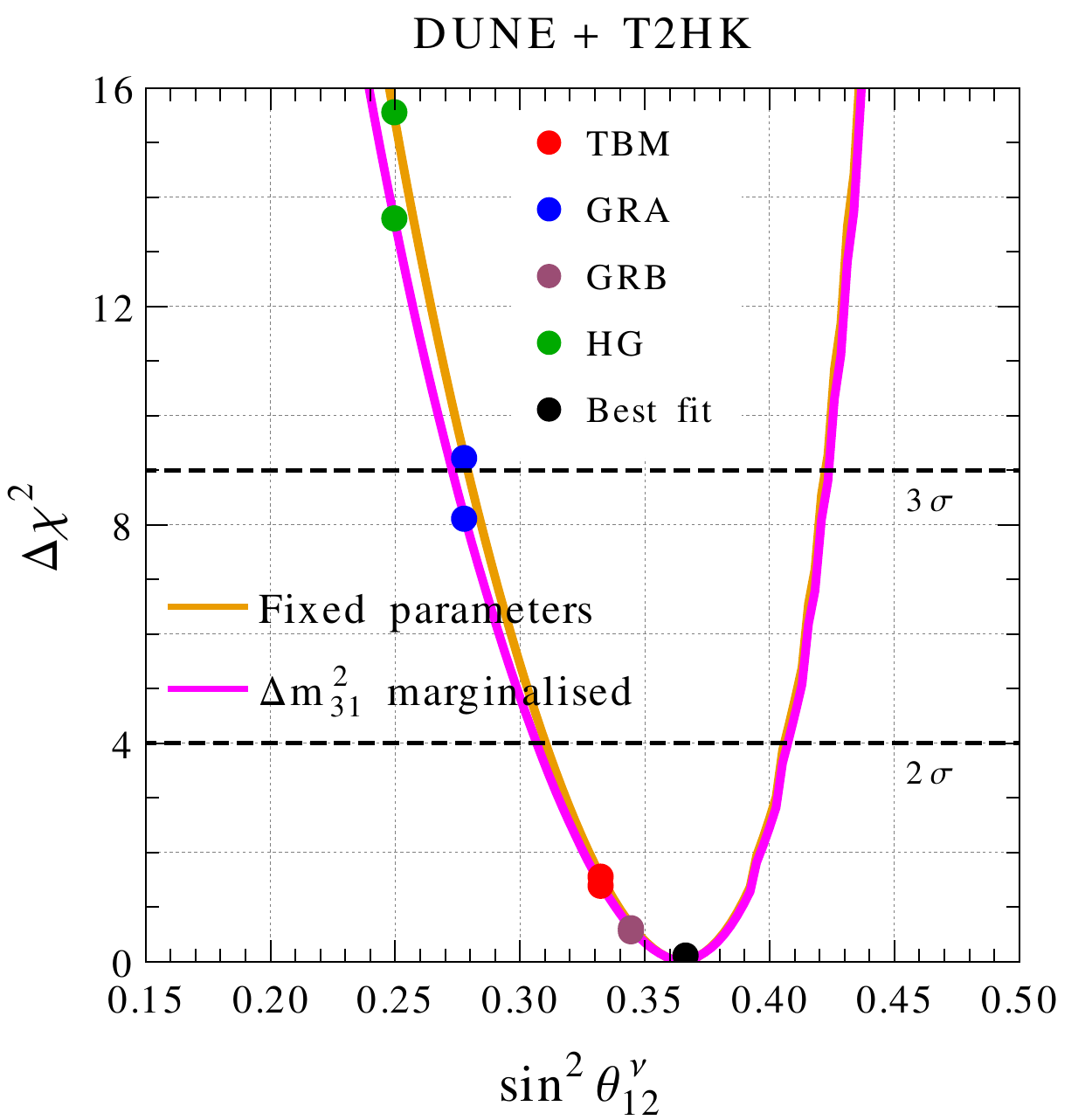}
\caption{Compatibility between various symmetry forms and present
best fit values of neutrino oscillation parameters in light of the prospective 
data from the combined DUNE\,+\,T2HK set-up. The black dot corresponds 
to $\sin^2\th^\nu_{12} = 0.364$ obtained from eq.~\eqref{eq:cosdelta} 
using the present best fit values of $\th_{12}$, $\th_{13}$, $\th_{23}$, and $\dcp$. 
The coloured dots correspond to the values of $\sin^2\th^\nu_{12}$ for the 
TBM, GRA, GRB, and HG symmetry forms. We show the results for the two
different cases: (i) fixed parameter scenario where all the oscillation 
parameters are kept fixed in the fit, and (ii) we only marginalise over 
$\Delta m_{31}^{2}$ in the fit in its present 3$\sigma$ allowed range.}
\label{fig:dm31sq-marginalisation}
\end{figure}

In this appendix, we give Fig.~\ref{fig:dm31sq-marginalisation} to
show the impact of the present 3$\sigma$ uncertainty on 
$\Delta m_{31}^{2}$ while testing the compatibility between 
the considered symmetry forms and present oscillation data. 
In Fig.~\ref{fig:dm31sq-marginalisation}, we show the potential
of the combined DUNE\,+\,T2HK set-up for the two different 
cases: (i) fixed parameter scenario where we keep all the 
oscillation parameters fixed at their benchmark values in 
the fit, and (ii) we only marginalise over $\Delta m_{31}^{2}$
in the fit in its present 3$\sigma$ allowed range. 
The fixed parameter curve is exactly similar to what we have 
already presented in Fig.~\ref{fig:currentBFV} for the combined
DUNE\,+\,T2HK set-up. For the GRB and TBM schemes,
we do not see much difference between the fixed parameter
case and the case where we marginalise over $\Delta m_{31}^{2}$
in the fit. For the GRA (HG) symmetry form, $\Delta\chi^2$
gets reduced by $\sim$ 11\% (13\%), when we marginalise
over $\Delta m_{31}^{2}$ instead of keeping it fixed in the fit.

\section{Agreement between Various Mixing Schemes 
and Oscillation Data in $\bo{(\sin^2\th_{23}^{\rm true},\dcp^{\rm true})}$ Plane}
\label{app:ss23delta}

\begin{figure}
\centering
\includegraphics[width=\textwidth]{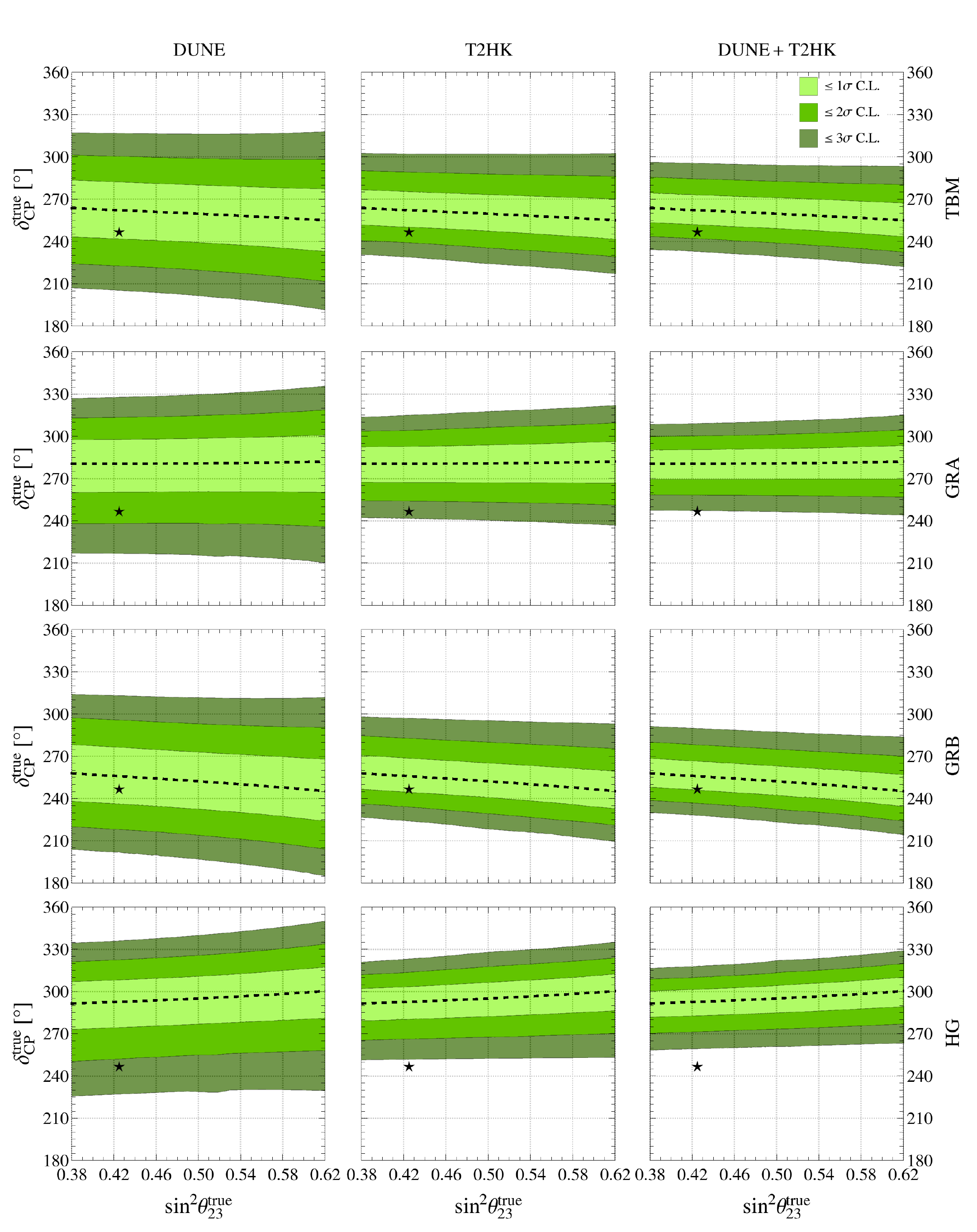}
\caption{Compatibility of various symmetry forms 
with any potentially true values of $\sin^2\th_{23}$ and $\dcp$ 
in the context of DUNE (left panels), T2HK (middle panels), 
and DUNE\,+\,T2HK (right panels).
For a given symmetry form (fixed $\th^\nu_{12}$), the black dashed line 
has been obtained using eq.~\eqref{eq:cosdelta} and fixing $\th_{12}$ and $\th_{13}$ 
to their NO best fit values. The star indicates the present best fit values 
for NO as given in Table~\ref{tab:oscparams}.
For all the symmetry forms, a significant part of the parameter space 
gets disfavoured at more than $3\sigma$ C.L. for DUNE\,+\,T2HK.}
\label{fig:th23delta}
\end{figure}

 Here we will see which regions of the parameter space in the plane of true values of 
$\sin^2\th_{23}$ and $\dcp$ will be compatible at less than $3\sigma$ C.L. 
for each symmetry form of interest, if that form is realised in Nature. 
To this aim, for each symmetry form (fixed $\th^\nu_{12}$), 
we calculate $\dcp$ using eq.~\eqref{eq:cosdelta} with the test 
values of the mixing angles $\th_{ij}^{\rm test}$. 
Then, we marginalise $\Delta\chi^2$ over $\th_{ij}^{\rm test}$ for given true values 
$\th_{23}^{\rm true}$ and $\dcp^{\rm true}$. 
The coloured bands in Fig.~\ref{fig:th23delta} represent potentially true values of 
$\dcp$ as well as $\sin^2\th_{23}$ with which the form under consideration is compatible 
at $1\sigma$, $2\sigma$, $3\sigma$ confidence levels in the context of 
DUNE, T2HK and their combination. 
If the true value of $\dcp$ turned out to lie outside 
these bands, this would imply that the given symmetry form is disfavoured 
at more than $3\sigma$ C.L. 
For all the symmetry forms, 
a significant part of the parameter space gets disfavoured at more than $3\sigma$.  

 For each symmetry form, the black dashed line
has been obtained using eq.~\eqref{eq:cosdelta} with $\th_{12}$ and $\th_{13}$ 
fixed to their NO best fit values. 
Note that a given symmetry form is 
well compatible with any point close to this line. 
The star denotes the present best fit values of 
$\sin^2\th_{23}$ and $\dcp$  
for the NO spectrum as given in Table~\ref{tab:oscparams}, 
namely, $(\sin^2\th_{23},\dcp) = (0.425,248\degree)$. 
As we can see from the right panels of Fig.~\ref{fig:th23delta}, 
i.e., in the context of DUNE\,+\,T2HK, 
the HG (GRA) form is disfavoured at more than (precisely) $3\sigma$~C.L., 
while the GRB and TBM forms are compatible with the star at $1\sigma$ and $2\sigma$, 
respectively. If the star moves in the future to a different point, we will immediately conclude 
which symmetry forms are (dis)favoured. 
Let us assume, e.g., that the future best fit values are 
$(\sin^2\th_{23},\dcp) = (0.58,300\degree)$. 
Then, in the context of DUNE\,+\,T2HK, the GRB and TBM forms would be disfavoured 
at more than $3\sigma$, while the HG (GRA) symmetry form would be compatible with 
this hypothetical position of the star at $1\sigma$ ($2\sigma$) confidence level.

 Finally, we would like to notice the compatibility between this figure and the numbers 
in Table~\ref{tab:summary}. Let us consider an example, in which 
GRB is the true form and HG is tested against it. 
In this case, from Table~\ref{tab:summary}, we read the C.L. at which these two 
symmetry forms can be distinguished by DUNE\,+\,T2HK, namely, $3.3\sigma$. 
We recall that the results in this table have been obtained assuming 
the current best fit values of the mixing angles to be their true values. 
Thus, for GRB we have $(\sin^2\th_{23},\dcp) = (0.425,256\degree)$, which 
are the true values of these parameters in the case under consideration. 
Now, we put this point on the right bottom panel of Fig.~\ref{fig:th23delta}, 
corresponding to the HG symmetry form for DUNE\,+\,T2HK,  
and find that this point falls just outside the dark green band 
representing $\dcp$ values compatible with the HG form at $3\sigma$ C.L.
It means that if $\dcp$ predicted by GRB together with the present  
best fit values of $\th_{23}$, $\th_{12}$, and $\th_{13}$ 
are realised in Nature, then the HG symmetry form will be disfavoured by 
DUNE\,+\,T2HK at $>3\sigma$. 
The same message is conveyed from Table~\ref{tab:summary} as well.

\bibliography{DUNEandT2HKv2}

\end{document}